\documentclass[pre,twocolumn,showpacs,superscriptaddress,preprintnumbers,amsmath,amssymb,floatfix]{revtex4-1}

\usepackage{graphicx}
\usepackage{dcolumn}
\usepackage{bm}
\usepackage{color}
\usepackage{ulem}
\usepackage{array}
\usepackage{stfloats}
\usepackage{braket}

\newcolumntype{P}[1]{>{\centering\arraybackslash}p{#1}}

\begin{document}

\preprint{APS/123-QED}

\title{Contribution of electron-atom collisions to the plasma conductivity of noble gases}

\author{S.~Rosmej}
\affiliation{Institut f\"ur Physik, Universit\"at Rostock, 18051 Rostock, Germany}
\author{H.~Reinholz}
\affiliation{Institut f\"ur Physik, Universit\"at Rostock, 18051 Rostock, Germany}
\affiliation{University of Western Australia, WA 6009 Crawley, Australia}
\author{G.~R\"opke}
\affiliation{Institut f\"ur Physik, Universit\"at Rostock, 18051 Rostock, Germany}

\date{\today}

\begin{abstract}
We present an approach which allows the consistent  treatment of bound states in the context of the dc conductivity in dense partially ionized noble gas plasmas. Besides electron-ion and electron-electron collisions, further collision mechanisms owing to neutral constituents are taken into account. Especially at low temperatures $T\approx 1 {\rm eV}$, electron-atom collisions give a substantial contribution to the relevant correlation functions.  We suggest an optical potential for the description of the electron-atom scattering which is applicable for all noble gases. The electron-atom momentum-transfer cross section is in agreement with experimental scattering data. In addition the influence of the medium is analysed, the optical potential is advanced including screening effects. The position of the Ramsauer minimum is influenced by the plasma. Alternative approaches for the electron-atom potential are discussed. Calculations of the electrical conductivity are compared with experimental data. 
\end{abstract}



\maketitle

	\section{Introduction \label{Intro}}
	
Properties of strongly coupled plasmas are of high interest for laser-induced dense plasmas (warm dense matter, WDM) and astrophysical systems. Up to now, the description of such partially ionized plasmas (PIP) is  still a challenge in plasma physics because the quantum-statistical treatment of electron-atom ($e$--$a$) collisions in dense matter is complex. Transport properties are an important input for simulations.
In this work, the electrical conductivity is investigated for dense noble gases at temperatures of $T \approx 1 {\rm eV}$ where the plasma is non-ideal and partially ionized.

Fully ionized plasmas (FIP) are analysed by the kinetic theory approach. The Boltzmann equation is solved e.g. by Spitzer \cite{Spitzer53} using a Fokker-Planck equation, Brooks-Herring \cite{Brooks} or Ziman \cite{Ziman} using the relaxation time approximation, commonly known analytical formulas for the electrical conductivity are given. 
The electron-ion ($e$--$i$) collisions are well described in a wide parameter range, whereas the treatment of $e$--$a$ collisions is not trivial within an analytical approach.
Within the Linear-Response-Theory (LRT) the treatment and inclusion of different collision mechanisms is vivid. In recent years, the influence of electron-electron ($e$--$e$) collisions at arbitrary degeneracy was discussed in Born approximation for a fully ionized plasma (FIP) by Reinholz {\it et al.} \cite{R4} and Karakhtanov \cite{Karakhtanov16}, leading to a systematic interpolation between the Spitzer and the Ziman limit, see also Esser {\it et al.} \cite{Esser98}. A correction factor is obtained which depends in contrast to the Spitzer value on density and temperature. 

Nevertheless these formulas valid for a FIP are not sufficient for the description of dense partially ionized noble gases, because bound states play an important role. Their treatment in a quantum-statistical approach is possible if considering a chemical picture. Here, the electron-ion bound-states are identified as atomic species. We consider a PIP containing atoms, electrons and singly charged ions ($n_{\rm e}=n_{\rm i}$) with a heavy particle density $n_{\rm heavy} = n_{\rm a}+n_{\rm i}$. For the considered densities multiple ionized ions are not relevant, a generalization is possible. Results for hydrogen are analyzed in \cite{Roepke89,Reinholz95,Redmer97}. The scattering mechanism between electrons and atoms in the correlation functions 
are immediatly included. The electron-atom ($e$--$a$) collisions are in particular relevant for non-degenerate plasmas if the system is below the Mott density. Above the Mott density the bound-states are dissolved. 

For the analysis of $e$--$a$ collisions, two important points have to be solved. In the chemical picture, the $e$--$a$ correlation function depends on the composition of the plasma, which can be determined from the mass-action laws. The Saha equations have been studied for a long time by many auhors, e.g. Ebeling {\it et al.}, see \cite{Ebeling82,Ebeling88} and F\"orster \cite{Foerster92} using Pad\'e techniques. The program package COMPTRA, for details see \cite{Kuhlbrodt04}, allows the calculation of  composition and transport properties in nonideal plasmas. It has been successfully applied to metal plasmas, see also \cite{Kuhlbrodt00,Kuhlbrodt01}.

In contrast to this, experimental data for noble gases have been understood only qualitatively, see \cite{Kuhlbrodt05}. A temperature dependent minimum in the conductivity can be observed as a consequence of the composition, but discrepancies with the experiment can be up to two orders of magnitude.
Kuhlbrodt {\it et al.} \cite{Kuhlbrodt05} believes that this results from using the polarization potential for the $e$--$a$ collisions. In \cite{Adams07}, Adams {\it et al.} verified this statement comparing the calculated momentum-transer cross sections using a polarization potential with measured data. Subsequently, using the experimental data obtained for the isolated $e$--$a$ momentum-transfer cross section and the composition calculated by COMPTRA04, Adams obtained quantitatively good agreement with experimental results for the conductivity of noble gases. 
Recently, results from swarm-derived cross sections using different Boltzmann solvers and Monte Carlo simulations 
have been obtained and compared with experimental data for the isolated $e$--$a$ scattering process, see \cite{SimuAr,SimuHeNe,SimuKrXe}. Especially for low energies, a good agreement has been found.

However, plasma effects like screening cannot be included by using the experimental transport cross sections for isolated systems. 
An alternative to the polarization potential is the construction of a so-called  optical potential \cite{Vanderpoorten1975,Paikeday76,SurGosh1982,Pangantiwar1989}. Such a  potential model has been successfully  applied to describing the $e$--$a$ momentum-transfer cross sections of all noble gases in a wide energy region, see \cite{Adibzadeh05}. The downside is the large number of free parameters and the fact, that no unique expressions
for all noble gases is known. In this work, we show that a modification in the approximation of the local exchange term leads to a universal expression for the cross sections for all noble gases. Adapting only one free parameter we find  good agreement with the experimental data.

In order to take into account the plasma environment, the optical potential for isolated systems is further modified by a systematic treatment of statical screening. Specific screening effects on the $e$--$a$ potential for a hydrogen plasma in the second order of perturbation, neglecting the exchange effects, have been discussed by Karakhtanov \cite{Karakhtanov2006}. Similar characteristics, e.g. a repulsive $e$--$a$ potential at large distances,
 are  found for the noble gases in this work.

\section{Electrical conductivity in T matrix approximation}
\label{sec:GMA}
Within the LRT transport coefficients of Coulomb systems are obtained by equilibrium correlation functions. According to the Zubarev formalism \cite{ZMR2,Roepke13} the static electrical conductivity is represented by
\begin{align} \label{eq:cond}
\sigma^{(L)} = -\frac{e^2\beta}{m^2 \Omega} \begin{vmatrix} 0 & N_{11} & \cdots & N_{1L} \\ N_{11} & d_{11} & \cdots & d_{1L} \\ \vdots & \vdots & \ddots & \vdots \\ N_{L1} & d_{L1} & \cdots & d_{LL} \end{vmatrix} / \begin{vmatrix} d_{11} & \cdots & d_{1L} \\ \vdots & \ddots & \vdots \\ d_{L1} & \cdots & d_{LL} \end{vmatrix} \, .
\end{align}
$N_{ll'}=\frac{1}{3}(\textbf{P}_l,\textbf{P}_{l'})$ is the Kubo scalar product
\begin{align}
 N_{ll'} &= N_{l'l} = \frac{n_{\rm e} \Omega_{\rm N} m}{\beta} \frac{\Gamma(\frac{l+l'+3}{2})}{\Gamma(\frac{5}{2})} \frac{I_{\frac{l+l'-1}{2}}(\alpha)}{I_{\frac{1}{2}}(\alpha)} \, , \label{eq:Nll}
\end{align}
where $\Omega_{\rm N}$ is the normalization volume, $\Gamma(x)$ is the Gamma functions, $I_{\nu}(\alpha)$ are the Fermi integrals and $\alpha=\beta \mu_{\rm e}^{\rm id}$ is the degeneracy of electrons. The degeneracy $\alpha$ is related to the degeneracy parameter $\Theta=(3\pi^2 n_{\rm e})^{-2/3} (2mk_{\rm B}T/\hbar^2)$. The plasma is further characterized by the coupling parameter $\Gamma=(4\pi n_{\rm e}/3)^{1/3}e^2/(4\pi\epsilon_0 k_{\rm B}T)$.
Considering the dominant contributions by the electron current we use the generalized moments $\textbf{P}_l$ as relevant observables \cite{Roepke13}:
\begin{equation}
 \textbf{P}_l = \sum \limits_k \hbar \textbf{k}(\beta E_k)^{\frac{l-1}{2}}n_k \, ,
\end{equation}
with $\beta=(k_{\rm B}T)^{-1}$, $\hbar {\bf k}$ the moment of electrons and $E_k=\hbar^2k^2/2m$ the kinetic energy of the electrons.

The generalized force-force correlation functions $d_{ll'}=\frac{1}{3} \braket{\dot{\textbf{P}}_l; \dot{\textbf{P}}_{l'}}_{i\varepsilon}$ contain the generalized forces $\dot{\bf P}_l=i [H,{\bf P}_l]/\hbar=i [V,{\bf P}_l]/\hbar$. Separating the interaction potential into the $e$--$i$, $e$--$e$ and $e$--$a$ contributions ($V=V_{\rm ei}+V_{\rm ee}+V_{\rm ea}$), the generalized force-force correlation functions are splitted into the different scattering mechanisms
\begin{align}
 d_{ll'} &= d_{l'l} = d_{ll'}^{\rm ei} + d_{ll'}^{\rm ee} + d_{ll'}^{\rm ea} \, .
\end{align}
The contributions $d_{ll'}^{{\rm e}c}$ are calculated in T matrix approximation, see also \cite{Meister82,Redmer97,Roepke88}.
For electron-ion and electron-atom collisions we treat the ions and atoms classical. Within the adiabatic limit the correlation functions $d_{ll'}^{\rm ei}$ and $d_{ll'}^{\rm ea}$ can be simplified to ($c\neq {\rm e}$)
\begin{align}
 d_{ll'}^{{\rm e}c} &= \frac{\hbar^3 n_c \Omega_{\rm N}}{3\pi^2m} \int\limits_0^{\infty} dk \, k^5 (\beta E_k)^{\frac{l+l'}{2}-1} f_k(1- f_k) Q_{\rm T}^{{\rm e}c}[V_{{\rm e}c}] \, ,
\end{align}
with the Fermi distribution function for electrons $f_k$. $Q_{\rm T}^{{\rm e}c}[V_{{\rm e}c}]$ is the momentum-transfer cross section:
\begin{align}
 Q_{\rm T}^{{\rm e}c} &= 2\pi \int\limits_{-1}^1 d(\cos\vartheta) \, \left[ 1- \cos\vartheta \right] \left( \frac{d \sigma^{{\rm e}c}}{d \Omega} \right) \, .
\end{align}
The interaction between electrons and ions in the plasma is described by a dynamically screened Coulomb potential, see \cite{Roepke88}. In the adiabatic limit the dynamically screened Coulomb potential is replaced by a statically screened Coulomb potential (Debye potential)
\begin{align}
 V_{\rm ei}(r) &= V_{\rm D}(r) = -\frac{e^2}{4\pi \epsilon_0} \frac{e^{-\kappa r}}{r} \, ,
\end{align}
with the inverse screening length $\kappa^2=(2\Lambda_{\rm e}^{-3}I_{-1/2}(\alpha)+n_{\rm i})\beta e^2/\epsilon_0$ and the thermal wavelength $\Lambda_{\rm e}=(2\pi\beta\hbar^2/m)^{1/2}$.
A recent discussion of the ionic contribution on the correlation functions is given by Rosmej \cite{Rosmej16}.

The $e$--$e$ contribution $d_{ll'}^{\rm ee}$ is more complex, see \cite{Redmer97,R4}. In the region of partially ionized noble gases the free-electron density in the plasma is low ($\Theta>1$), the $e$--$e$ correlation function can be simplfied to
\begin{align}
 d_{33}^{{\rm ee}} &= \frac{8 n_{\rm e}^2 \Omega_{\rm N}}{3\sqrt{\pi}\beta} \sqrt{\frac{m}{\beta}} \int\limits_0^{\infty} dP \, P^7 e^{-P^2} Q_{\rm v}^{{\rm ee}}[V_{\rm ee}] \, .
\end{align}
$Q_{\rm v}^{{\rm ee}}[V_{\rm ee}]$ is the viscosity cross section:
\begin{align}
 Q_{\rm v}^{{\rm ee}} &= 2\pi \int\limits_{-1}^1 d(\cos\vartheta) \, \left[ 1- \cos^2\vartheta \right] \left( \frac{d \sigma^{{\rm ee}}}{d \Omega} \right) \, .
\end{align}
In general, the $e$--$e$ interaction is also described by a dynamically screened Coulomb potential. For partially ionized plasma the free electron density $n_{\rm e}$ is low and the replacement of the dynamical screening by a statical screening is applicable $V_{\rm ee}(r)=-V_{\rm D}(r)$, see \cite{Roepke88}. The effect of dynamical screening by an effective screening radius, see \cite{Roepke88}, is discussed in \cite{R4} which leads to a difference in the results by less than $2 \%$. Alternatively, dynamical screening can be taken into account by a Gould-DeWitt sheme, see \cite{GdW}.

The cross sections are calculated via the partial wave decomposition ($c={\rm i,a}$)
\begin{align}
 Q_{\rm T}^{{\rm e}c}(k) &= \frac{4\pi}{k^2} \sum_{\ell = 1}^{\infty} \ell \sin^2\left[ \delta_{\ell-1}^{{\rm e}c}(k) - \delta_{\ell}^{{\rm e}c}(k) \right] \, , \\
\begin{split}
 Q_{\rm v}^{\rm ee}(k) &= \frac{4\pi}{k^2} \sum_{\ell = 1}^{\infty} \left[ 1+ \frac{(-1)^{\ell}}{2}\right] \frac{\ell(\ell+1)}{2\ell +1} \times \\
& \qquad \qquad \quad \sin^2\left[ \delta_{\ell-1}^{\rm ee}(k) - \delta_{\ell+1}^{\rm ee}(k) \right] \, ,
\end{split}
\end{align}
with the scattering phase shifts $\delta_{\ell}$. The phase shift calculations are performed using Numerov's method.
The convergence of these cross sections depends on the temperatures. With increasing temperature the number of relevant phase shifts increases. For high temperatures $T>10^6 {\rm K}$ the Born approximation (BA) is applicable:
\begin{align} \label{eq:dqdwBorn}
 \left(\frac{d\sigma^{{\rm e}c}}{d\Omega}\right)_{\rm BA} &= \frac{\mu^2_{{\rm e}c}\Omega^2_{\rm N}}{4 \pi^2 \hbar^4}  | \tilde V_s(q) |^2 \left( 1 - \frac{\delta_{{\rm e}c}}{2} \left| \frac{\tilde V_s(q')}{\tilde V_s(q)} \right| \right) \, , 
\end{align}
with the Fourier transformed potential $\tilde V(q)=\Omega_{\rm N}^{-1} \int d^3{\bf r} \, e^{i{\bf qr}} V(r)$, the relative mass $\mu_{{\rm e}c}= m_{\rm e}m_c/(m_{\rm e}+m_c)$, the transferred momenta $q={\bf |q|=|k-k'|}=2k\sin(\vartheta/2)$ and $q'={\bf |q'|=|k+k'|}=2k\cos(\vartheta/2)$.

\section{Electron-atom interaction in dense plasma}

\subsection{Optical potential for isolated systems}

Within the Green functions technique \cite{RRZ87} the interaction between electrons and atoms is expanded up to the second order of perturbation, see also \cite{Karakhtanov2006}
\begin{align} \label{eq:2ndOrderP}
 V_{\rm ea} &= V^{(1)} + V^{(2)} + V_{\rm ex} \, .
\end{align}
In \cite{RRZ87,Karakhtanov2006} the non-local exchange part $V_{\rm ex}$ is neglected. In this work we consider the role of exchange in a local approximation, see Sec.~\ref{sec:ex}.

The first order term $V^{(1)}$ describes the Coulomb interaction between the free electron with the nucleus as well as with the bound electrons. This part is commonly known as the Hartree-Fock potential $V^{(1)} \equiv V_{\rm HF}(r)$ given by
\begin{align} \label{eq:VHF}
 V_{\rm HF}(r) &= -\frac{e^2}{4\pi \epsilon_0 r} \int\limits_r^{\infty} dr' \, 4\pi r' \rho(r') \left( r' - r \right) \, ,
\end{align}
where $\rho(r)$ is the bound electron density in the target atom, which is here calculated taking the atomic Roothaan-Hartree-Fock wave functions \cite{Clementi74}. For large distances 
$V_{\rm HF}(r)$ decreases exponentially.

In contrast to this result the large distances behavior for the interaction between electrons and neutral particles is obtained by Born and Heisenberg \cite{BornHeisenberg1924} as $V_{\rm ea}(r \rightarrow \infty) \propto -r^{-4}$. 
Using the Green functions technique Redmer {\it et al.} found this behavior by the second order of perturbation which is related to the polarization potential $V^{(2)} \equiv V_{\rm P}$, see \cite{RRZ87}.
For the polarization potential we take the form used by Paikeday \cite{Paikeday2000}:
\begin{align} \label{eq:VPP}
 V_{\rm P}(r) &= -\frac{e^2 \alpha}{8\pi \epsilon_0 (r+r_0)^4} \, ,
\end{align}
where $\alpha$ is the dipole polarizability and $r_0$ is a cut-off parameter which we take here as a free parameter in the scale of the Bohr radius $a_0$. 

In general the polarization potential depends on the energy. However, Paikeday obtained a weak dependence on the energy, which is negligible for our considerations, see \cite{Paikeday2000}. In contrast to the form Eq.~(\ref{eq:VPP}) a lot of different analytical forms for the polarization potential are used, see \cite{Paikeday76,RRZ87,Bottcher71,Schrader79}. The cut-off parameter $r_0$ is also treated in a different way. In \cite{RRZ87} this parameter is taken to give the correct value for $V_{\rm P}(0)$. Mittleman and Watson \cite{Mittleman59} derived an analytical formula considering semi-classical electrons $r_0=(\alpha a_0/(2Z^{1/3}))^{1/4}$. Nevertheless for short distances the finite polarization part is negligible in contrast to the dominant divgerent Coulomb interaction in the Hartree-Fock term. So it seems to be desirable to choose $r_0$ optimal for intermediate distances in contrast to the short distances case $r=0$. Alternatively, the cut-off parameter is adjusted by the experimental data for the differential cross section for electron-atom collisions by Paikeday at intermediate energies, see \cite{Paikeday2000}. For positron-atom collisions (no exchange part), Schrader adjusted the cut-off parameter by experimental data for the scattering length, see \cite{Schrader79}. In this work the cut-off parameter is determined by a correct low energy behavior for the ligther elements helium and neon and by a correct description of the position and hight of the Ramsauer minimum for the havier elements argon, krypton and xenon. In Sec.~\ref{sec:QT}, the results are compared with the experimental data for the momentum-tranfer cross section. The values for the cut-off parameter $r_0$ are given in Tab.~\ref{tab:r0}.

\begin{center}
\begin{table}[h]
 \begin{tabular}{ | c | c c c c c |}
\hline
$r_0 [a_0]$ & He & Ne & Ar & Kr & Xe  \\ \hline
present work & 1.00 & 1.00 & 0.86 & 0.92 & 1.00 \\
Mittleman \cite{Mittleman59} & 0.86 & 0.89 & 1.21 & 1.27 & 1.38  \\
Paikeday \cite{Paikeday2000} & 0.92 & 1.00 & 2.89 & 3.40 & - \\
Schrader \cite{Schrader79} & 1.77 & 1.90 & 2.23 & 2.37 & 2.54 \\
\hline
 \end{tabular}
 \caption{Cut-off parameter $r_0$ in units of $a_0$.}
 \label{tab:r0}
\end{table}
\end{center}

With Eqs.~(\ref{eq:VHF},\ref{eq:VPP}) the electron-atom interaction Eq.~(\ref{eq:2ndOrderP}) is refered as the optical potential \cite{Vanderpoorten1975,Pangantiwar1989}
\begin{align} \label{eq:Vopt}
 V_{\rm opt}(r) &= V_{\rm HF}(r) + V_{\rm P}(r) + V_{\rm ex}(r) \, ,
\end{align}
where the exchange contribution is approximated by a local field.


\subsection{Approximation of the exchange potential} \label{sec:ex}
In the optical potential model the exchange contribution is considered in a local field approximation.
The general accepted exchange potential is derived in a semiclassical approximation (SCA) \cite{Furness1973,RileyTruhlar1976,Yau78}:
\begin{align}
\begin{split}
 V_{\rm ex}^{\rm SCA}(r,K_{\rm RT}(r)) &= \frac{\hbar^2}{4m} \left\{ K_{\rm RT}^2(r) \right. \\
& \quad \left. - \sqrt{K_{\rm RT}^4(r) + \frac{4me^2}{\hbar^2 \epsilon_0} \rho(r) } \right\} \, ,
\end{split}
\end{align}
with the local electron-momentum $K(r)$ in the introduced version of Riley and Truhlar \cite{RileyTruhlar1976, Yau78}
\begin{align} \label{eq:KRT}
 K_{\rm RT}^2(r) &= k^2+ \frac{2m}{\hbar^2} [|V_{\rm HF}(r)|+|V_{\rm P}(r)|] \, .
\end{align}

In competition to the SCA for a free-electron gas Mittleman and Watson \cite{Mittleman60} derived a local exchange potential, see also \cite{RileyTruhlar1976}
\begin{align} \label{eq:Vex}
 V_{\rm ex}^{\rm M}(r,K(r)) &= - \frac{e^2}{4\pi \epsilon_0} \frac{2}{\pi} K_{\rm F}(r) F(K(r)/K_{\rm F}(r)) \, ,
\end{align}
with the Fermi momentum $K_{\rm F}(r)= \left[3\pi^2 \rho(r)\right]^{1/3}$ and the function 
\begin{align}
 F(\eta) &= \frac{1}{2} + \frac{1-\eta^2}{4\eta} \ln\left| \frac{\eta+1}{\eta-1} \right|  \, .
\end{align}
The treatment of the local electron-momentum $K(r)$ is different. In contrast to Riley and Truhlar $K(r)=K_{\rm RT}(r)$, Eq.~(\ref{eq:KRT}), Hara assumed the inclusion of the ionization potential $I$ into the local momentum \cite{Hara1967}
\begin{align}
 K_{\rm H}^2(r) &= k^2+2 \frac{2m}{\hbar^2}I+K_{\rm F}^2(r) \, .
\end{align}
It is discussed by Hara \cite{Hara1967} as well as by Riley and Truhlar \cite{RileyTruhlar1976}, that $K_{\rm H}(r)$ leads to a wrong asymptotic behavior of the exchange potential for large distances. Some modifications are performed in \cite{SurGosh1982,RileyTruhlar1976}. 

In this work the local momentum of Riley and Truhlar is modified by
\begin{align} \label{eq:Kex}
 K_{\rm RRR}^2(r) &= k^2+ \frac{2m}{\hbar^2} [|V_{\rm HF}(r)|+|V_{\rm P}(r)|+|V_{\rm ex}^{\rm M}(r,0)|] \, .
\end{align}
In contrast to Riley-Truhlar's free-electron-gas exchange approximation (FER) we include the momentum-free exchange part $V_{\rm ex}^{\rm M}(r,0)=-\frac{e^2}{4\pi \epsilon_0}\frac{2}{\pi} K_{\rm F}(r)$ into the local momentum of Eq.~(\ref{eq:Vex}).

\begin{widetext}
\begin{center}
\begin{table}[h]
 \begin{tabular}{ | c | c c c c c c  c | c  c c c c c c | c  c c c c c c |}
\hline
Helium & & & & $\delta_0$ & &  &  & & &  &  $\delta_1$ & &  &  & & &  &  $\delta_2$ &  & &  \\
$ka_0$ & & (1) & (2) & (3) & (4) & (5) & &  & (1) & (2) & (3) & (4) & (5) & & & (1) & (2) & (3) & (4) & (5) & \\
\hline
0.10	&&2.994	&2.993	&3.006	&2.957	&3.337 & & 	&	&	&	&	& & &	&	&	&	&	& & \\
0.25	&&2.776	&2.770	&2.783	&2.691	&3.043 & &	&	&	&	&	& & & &	&	&	&	& &\\
0.50	&&2.436	&2.412	&2.422	&2.304	&2.648 & &	&0.043	&0.045	&0.076	&0.023	&0.218 & & &	&	&	&	& &\\
0.75	&&2.139	&2.093	&2.111	&2.001	&2.332 & &	&0.110	&0.101	&0.146	&0.064	&0.316 & & 	&0.005	&0.006	&0.009	&0.004	&0.013 &\\
1.00	&&1.890	&1.835	&1.856	&1.769	&2.071 & &	&0.183	&0.159	&0.205	&0.116	&0.359 & & 	&0.014	&0.015	&0.020	&0.010	&0.025 &\\
1.50	&&1.522	&1.473	&1.491	&1.446	&1.654 & &	&0.284	&0.247	&0.279	&0.212	&0.367 & & 	&0.042	&0.041	&0.047	&0.035	&0.053 &\\
\hline \hline
Neon & & & & $\delta_0$ & &  &  & & &  &  $\delta_1$ & &  &  & & &  &  $\delta_2$ &  & &  \\
$ka_0$ & & (1) & (2) & (3) & (4) & (5) & &  & (1) & (2) & (3) & (4) & (5) & & & (1) & (2) & (3) & (4) & (5) & \\
\hline
0.20	&&6.072	&6.104	&6.179	&6.028	&7.716 & & 	&	&	&	&	& & &	&	&	&	&	& & \\
0.30	&&5.965	&6.004	&6.069	&5.902	&7.056 & &	&	&	&	&	& & & &	&	&	&	& &\\
0.40	&&5.857	&5.899	&5.947	&5.779	&6.648 & &	&	&	&	&	& & & &	&	&	&	& &\\
0.50	&&5.748	&5.789	&5.820	&5.659	&6.361 & &	&3.040	&3.052	&3.030	&2.917	&3.196 & & 	&0.004	&0.005	&0.008	&0.002	&0.016 & \\
0.70	&&	&	&	&	& & &	&2.933	&2.949	&2.909	&2.785	&3.124 & & 	&	&	&	&	& &\\
0.80	&&	&	&	&	& & &	&2.873	&2.890	&2.845	&2.724	&3.073 & & 	&	&	&	&	& &\\
0.90	&&5.321	&5.349	&5.333	&5.215	&5.626 & &	&2.812	&2.828	&2.781	&2.667	&3.017 & & 	&	&	&	&	& &\\
1.00	&&5.219	&5.243	&5.222	&5.114	&5.480 & &	&2.751	&2.766	&2.719	&2.615	&2.958 & & 	&0.065	&0.059	&0.075	&0.034	&0.127 &\\
\hline \hline
Argon & & & & $\delta_0$ & &  &  & & &  &  $\delta_1$ & &  &  & & &  &  $\delta_2$ &  & &  \\
$ka_0$ & & (1) & (2) & (3) & (4) & (5) & &  & (1) & (2) & (3) & (4) & (5) & & & (1) & (2) & (3) & (4) & (5) & \\
\hline
0.10	&&9.274	&9.266	&9.374	&9.249	&11.858 & & 	&6.279	&6.276	&6.279	&6.275	&6.318 & &	&	&	&	&	& & \\
0.25	&&9.045	&9.015	&9.141	&8.987	&10.628 & &	&6.227	&6.193	&6.213	&6.192	&6.381 & & 	&0.002	&0.002	&0.004	&0.001	&0.031 &\\
0.50	&&8.647	&8.580	&8.658	&8.561	&9.250 & &	&6.001	&5.901	&5.938	&5.909	&6.214 & & 	&0.045	&0.035	&0.061	&0.024	&0.815 &\\
0.75	&&8.249	&8.164	&8.206	&8.162	&8.536 & &	&5.702	&5.583	&5.613	&5.604	&5.901 & & 	&0.277	&0.185	&0.284	&0.150	&1.963 &\\
1.00	&&7.875	&7.790	&7.808	&7.798	&8.030 & &	&5.411	&5.305	&5.320	&5.333	&5.575 & & 	&0.860	&0.581	&0.782	&0.536	&2.008 &\\
1.50	&&7.252	&7.170	&7.163	&7.182	&7.288 & &	&4.923	&4.861	&4.854	&4.892	&4.996 & & 	&1.644	&1.539	&1.614	&1.594	&1.877 &\\
\hline
 \end{tabular}
 \caption{e-He, e-Ne and e-Ar partial wave phase shifts $\delta_{\ell}$ (in rad) for $\ell=0,1,2$ omitting polarization part $V_{\rm P}=0$. \\ 
(1): static-exchange approximation (SEA); \\ 
(2): present work (RRR); \\ 
(3): semiclassical exchange approximation (SCA); \\ 
(4): Hara's free-electron-gas exchange approximation (FEH); \\ 
(5): Riley-Truhlar's free-electron-gas exchange approximation (FER).}
\label{tab:PSex}
\end{table}
\end{center}
\end{widetext}

The quality of the proposed local exchange approximations is verified due to a comparison of the scattering phase shifts with the static-exchange approximation (SEA) using the non-local exchange term. The SEA is the solution of the scattering process in first order of perturbation neglecting polarization effects. For a consistent comparison we omit the polarization potential ($V_{\rm PP}=0$) in this subsection.

In Tab.~\ref{tab:PSex} phase shifts calculations are performed for the different proposed local exchange potentials with the non-local exchange term performed for helium by Duxler {\it et al.} \cite{Duxler71}, for neon by Thompson \cite{Thompson66}, and for argon by Thompson again and by Pindzola and Kelly \cite{Pindzola74}. 
%
%
Hara's free-electron-gas exchange approximation underestimates the effects of exchange, wheras Riley-Truhlar's free-electron-gas exchange approximation overestimates the effects of exchange. For helium and argon this result was already obtained by Riley and Truhlar \cite{RileyTruhlar1976}.
The best agreement with the SEA phase shifts yield the popular SCE potential as well as the proposed exchange potential RRR Eq.~(\ref{eq:Vex},\ref{eq:Kex}). 
Both potentials leads to the same high-energy asymptotic limit.
For small wave numbers $\kappa a_0 \leq 0.5$ the proposed exchange potential leads to the best results with an error $\leq 0.1 \, {\rm rad}$.
In particular for helium and neon the proposed potential works very fine. The exchange part for both noble gases is more of interest as for argon because their polarizability is one order of magnitude lower as for argon. The polarization part for argon is of higher interest as for helium and neon. So for explicit calculations the momentum-transfer cross sections for helium and neon is more influenced by the exchange part as for argon.

\subsection{Plasma effects}
The dominant plasma effect is the screening of the electron-atom potential.
The screened optical potential is given by the screening of each contribution
\begin{align} \label{eq:sVopt}
 V_{\rm opt}^{\rm s}(r) &= V_{\rm HF}^{\rm s}(r) + V_{\rm P}^{\rm s}(r) + V_{\rm ex}^{\rm s}(r) \, .
\end{align}
The effect of screening was extensively analysed for the polarization potential and a screening factor $e^{-2\kappa r} (1+\kappa r)^2$ was obtained by Redmer and R\"opke \cite{RR85}
\begin{align} \label{eq:sVPP}
 V_{\rm P}^{\rm s}(r) &= V_{\rm P}(r) \, e^{-2\kappa r} (1+\kappa r)^2 \, .
\end{align}

The Hartree-Fock part for isolated systems is determined by the Coulomb interaction between the incoming electron with the core as well as with the shell electrons. 
Screening effects for this part can be considered by replacing the fundamental Coulomb interaction by a Debye potential. For partially ionized hydrogen plasma this is done by Karakhtanov \cite{Karakhtanov2006}. The screened Hartree-Fock potential is 
\begin{align} \label{eq:sVHF}
 V_{\rm HF}^{\rm s}(r) &= -\frac{Ze^{-\kappa r}}{r} + I_1 + I_2 + I_ 3 \, ,\\
I_1 &= \frac{1}{2\kappa r} e^{-\kappa r} \int\limits_0^{r} \frac{\rho(r_1)}{r_1} e^{\kappa r_1} \, dr_1 \, , \\
I_2 &= - \frac{1}{2\kappa r} e^{-\kappa r} \int\limits_0^{\infty} \frac{\rho(r_1)}{r_1} e^{-\kappa r_1} \, dr_1 \, ,  \\
I_3 &= \frac{1}{2\kappa r} e^{\kappa r} \int\limits_r^{\infty} \frac{\rho(r_1)}{r_1} e^{-\kappa r_1} \, dr_1 \, ,
\end{align}
which is derived in App.~\ref{AppA}. For large distances this potential becomes repulsive, see App.~\ref{AppB}.

In App.~\ref{AppB} the asymptotic behavior at large distances $r \rightarrow \infty$ for the screened Hartree-Fock part is derived for intermediate screening parameters $0<\kappa a_0<1$
\begin{align} \label{eq:asVHF}
 V_{\rm HF}^{\rm s}(r) &= \frac{Ze^2}{4\pi \epsilon_0} \frac{e^{-\kappa r}}{r} (\kappa a_0)^2 \, {\cal C}_0 \, , \\
 {\cal C}_0 &= \frac{Z^{-1}}{6} \int\limits_0^{\infty} \left( \frac{r_1}{a_0}\right)^2 4\pi r_1^2 \rho(r_1) \, dr_1 \, .
\end{align}
${\cal C}_0$ is a characteristic quantity for each element. For hydrogen we obtain ${\cal C}_0^{\rm H} = 1/2$ which is in agreement with the results in \cite{Karakhtanov2006}.
For the noble gases we find ${\cal C}_0^{\rm He} = 0.1974$, ${\cal C}_0^{\rm Ne} = 0.1563$, ${\cal C}_0^{\rm Ar} = 0.2411$, ${\cal C}_0^{\rm Kr} = 0.1830$ and ${\cal C}_0^{\rm Xe} = 0.1933$, see Tab.~\ref{tab:coef} in App.~\ref{AppB}.

For the screened exchange part we replace the Hartree-Fock and polarization part in Eq.~(\ref{eq:Kex}) by their screened ones Eq.~(\ref{eq:sVPP}) and Eq.~(\ref{eq:sVHF}).

\begin{figure}[htb] 
\begin{center}
 \includegraphics[width=0.99\linewidth,clip=true]{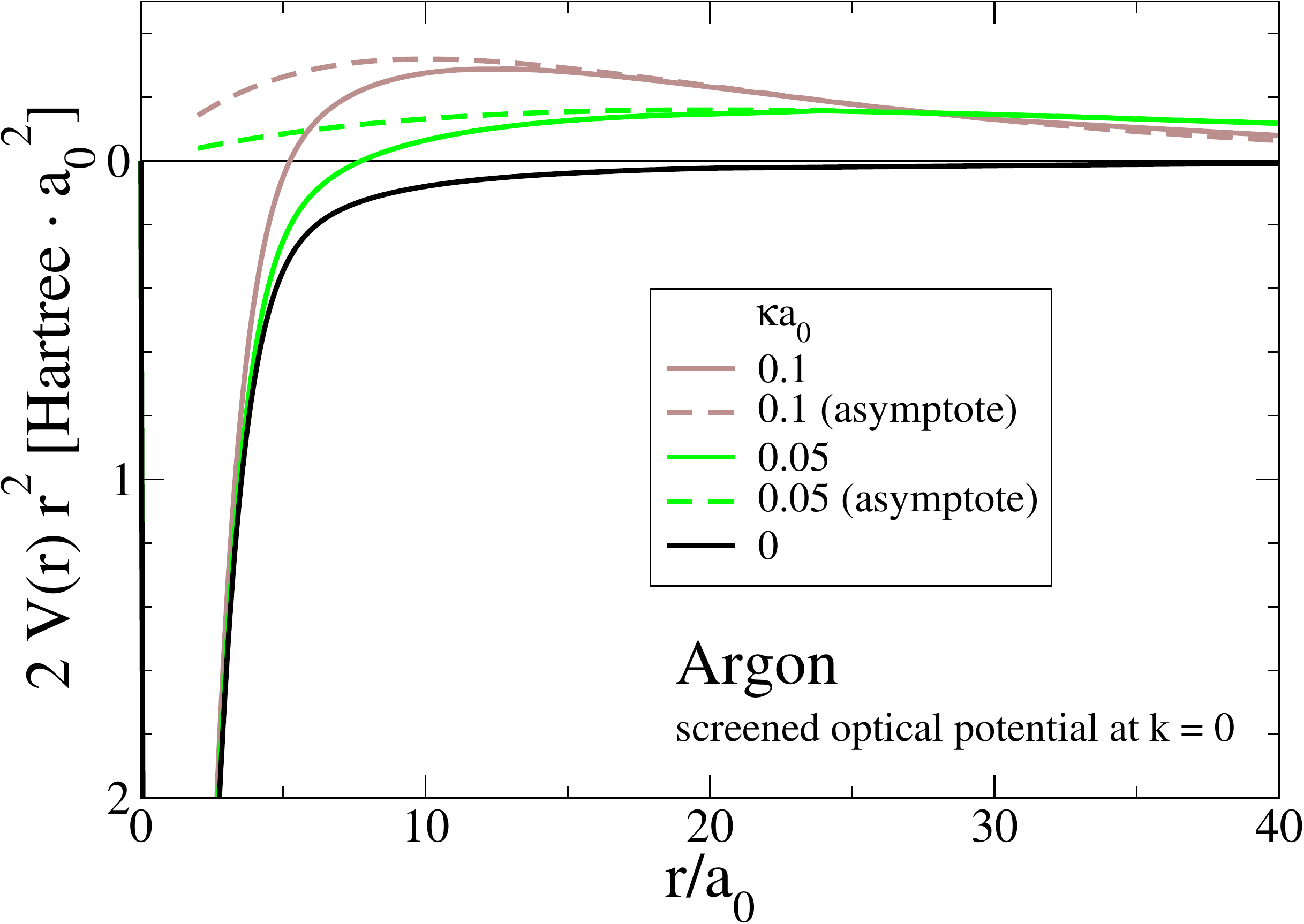}
\caption{(Color online) Electron-argon interaction described by the screened optical potential Eq.~(\ref{eq:sVopt}) for $\kappa a_0 =0.1$ and $\kappa a_0 = 0.05$ in comparison with the unscreened optical potential, Eq.~(\ref{eq:Vopt}). The asymptotes Eq.~(\ref{eq:asVHF}) are also shown.}
\label{fig:soptAr}
\end{center}
\end{figure}

%
%
%

In contrast to the unscreened case in which the polarization part is the main contribution to the optical potential at large distances we obtain the Hartree-Fock term is dominant, the screened optical potential become repulsive for all noble gases. In Fig.~\ref{fig:soptAr} this effect is illustrated for argon. Karakhtanov \cite{Karakhtanov2006} obtained the same effect for partially ionized hydrogen.

For short distances screening effects are not visible. We estimate for the momentum-transfer cross section small differences at higher energies. Great differences and qualitative different behaviors of the momentum-transfer cross sections are obtained at low energies. The results are given in Sec.~\ref{sec:sQT}.

\section{Results}

\subsection{Momentum-transfer cross section} \label{sec:QT}
The momentum-transfer cross section for the electron-atom interaction is determined by the phase shifts of the optical potential using the values for the cut-off parameter $r_0$ in Tab.~\ref{tab:r0}.

\begin{widetext}

\begin{figure}[htb] 
\begin{center}
 \includegraphics[width=0.4\linewidth,clip=true]{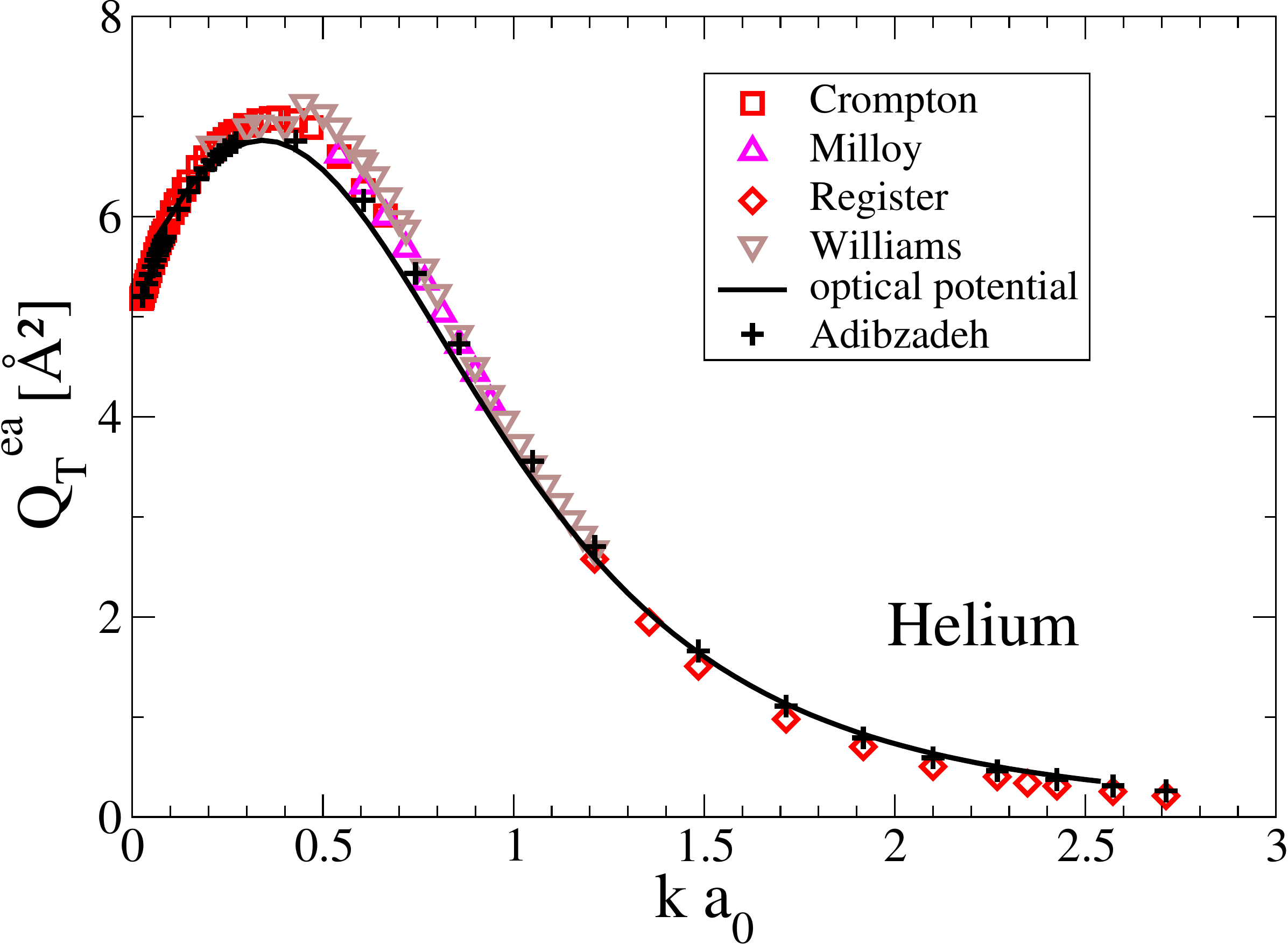}
 \includegraphics[width=0.4\linewidth,clip=true]{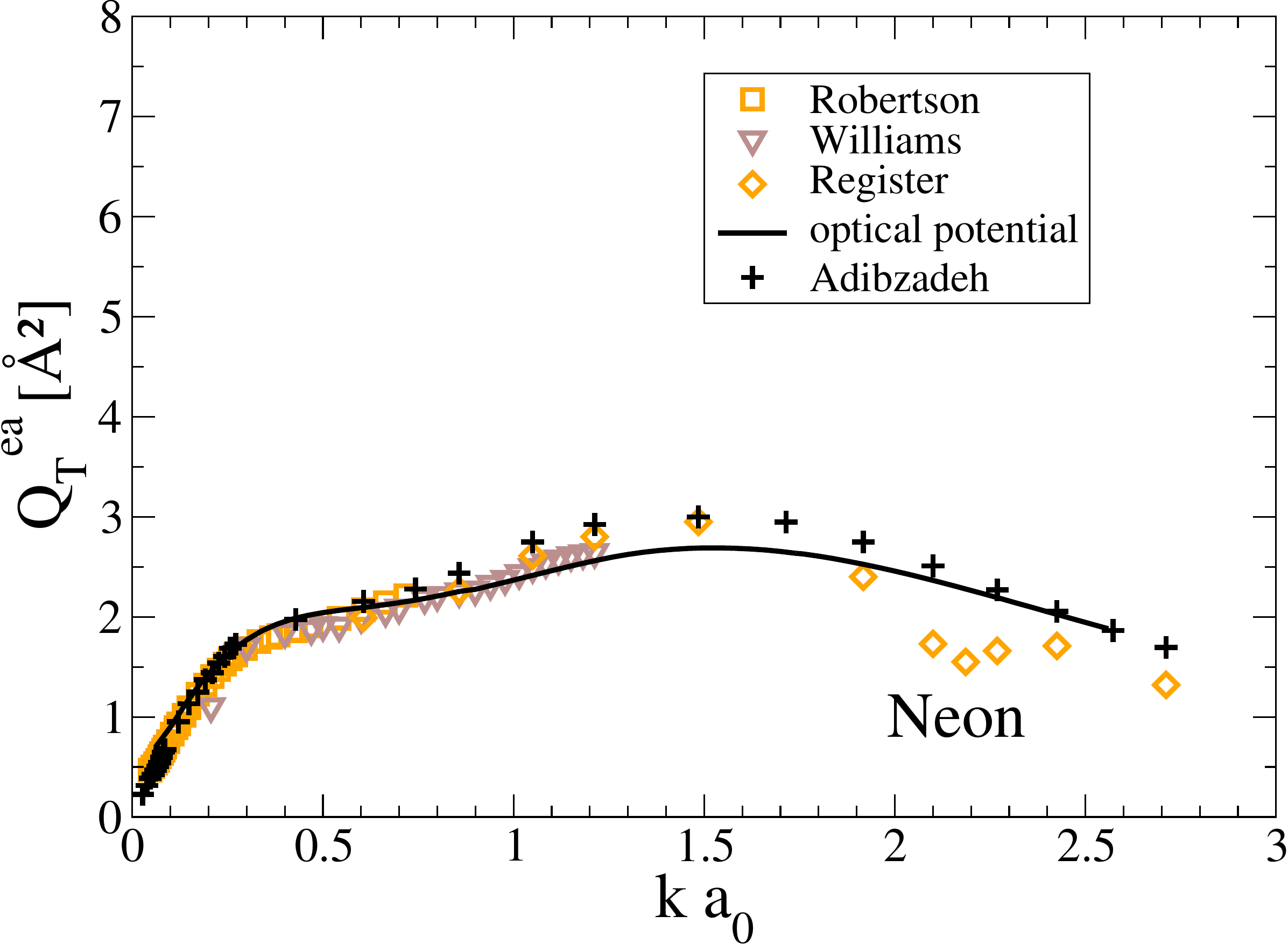}
 \includegraphics[width=0.4\linewidth,clip=true]{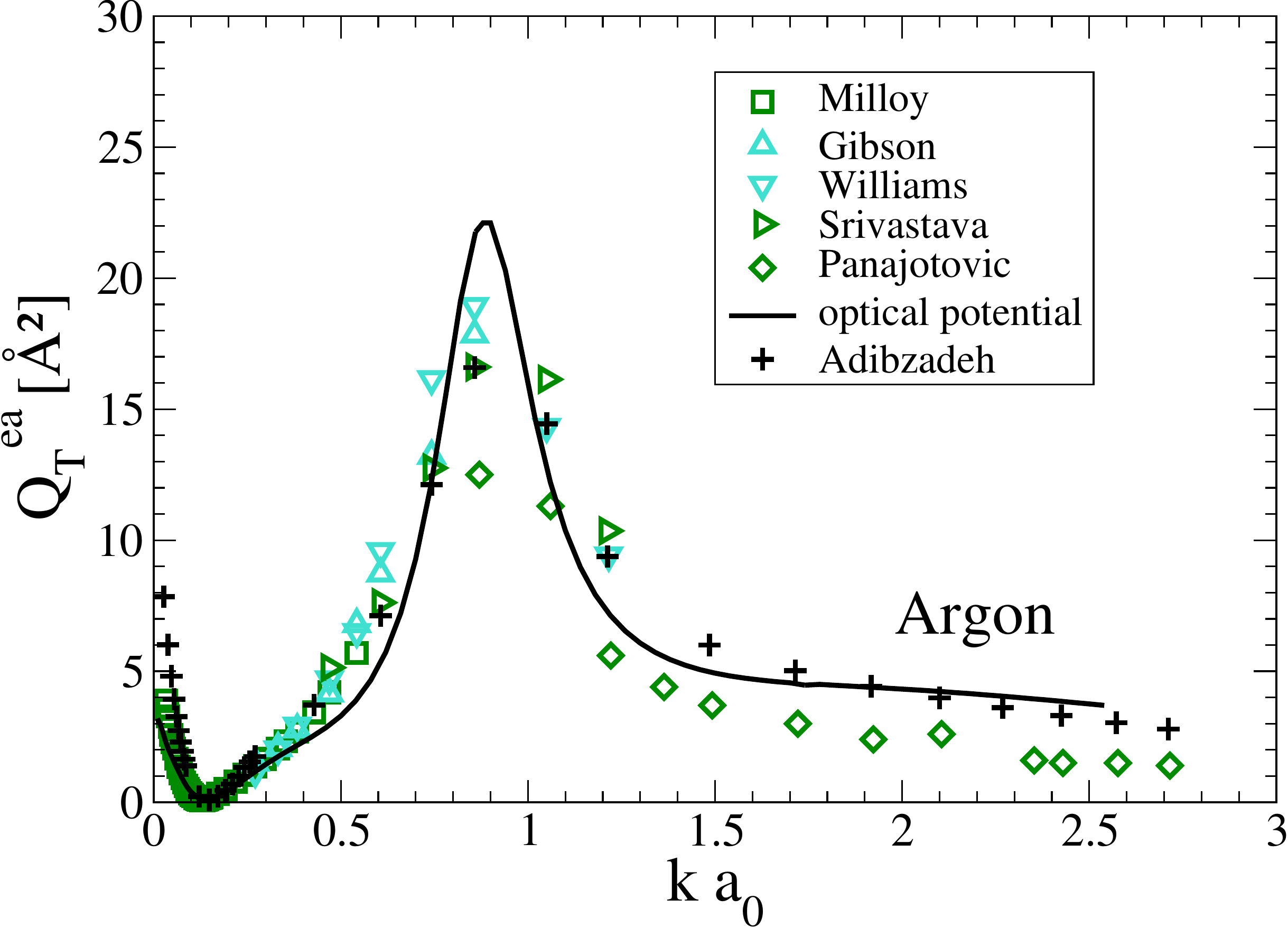}
 \includegraphics[width=0.4\linewidth,clip=true]{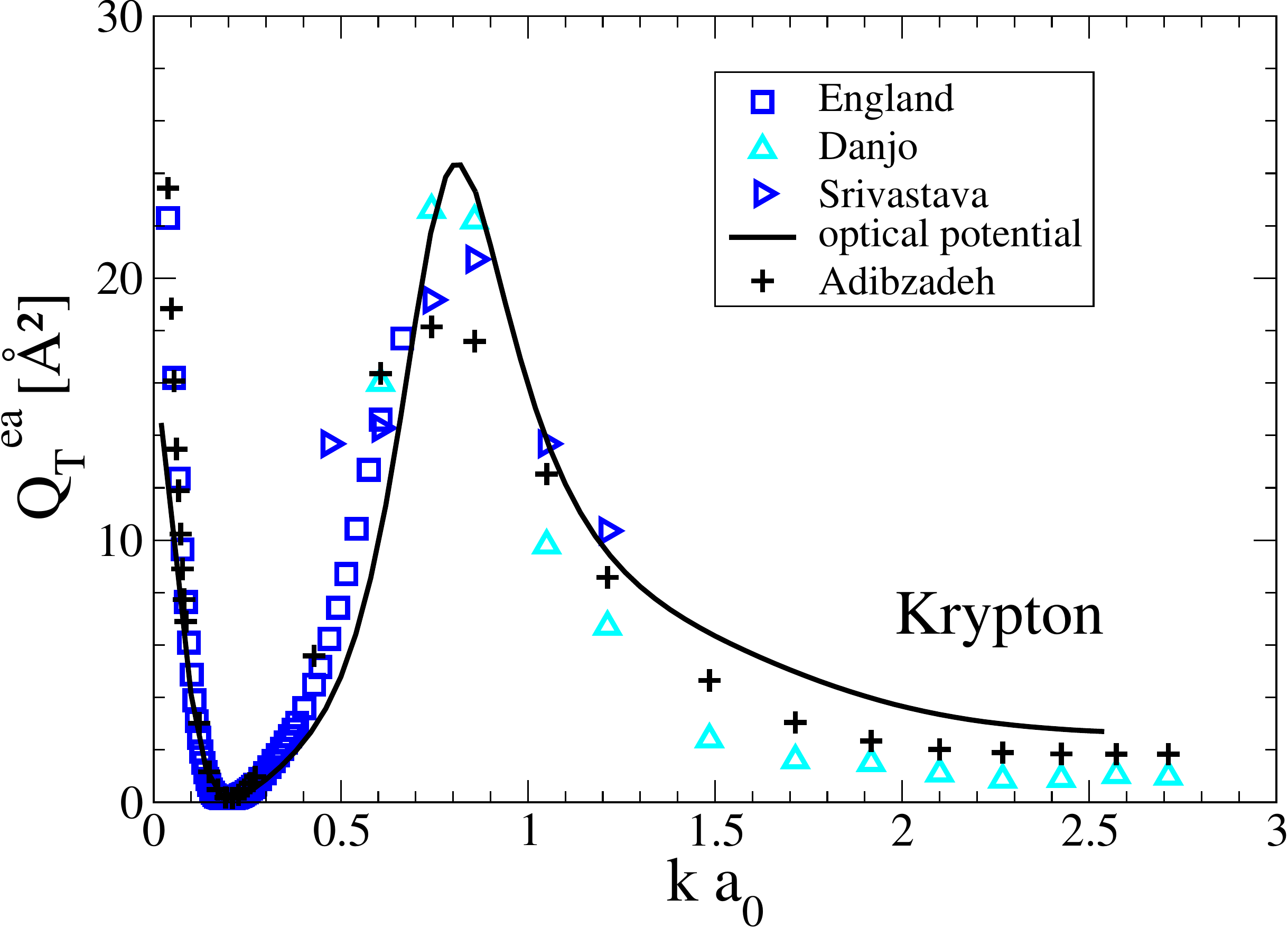}
 \includegraphics[width=0.4\linewidth,clip=true]{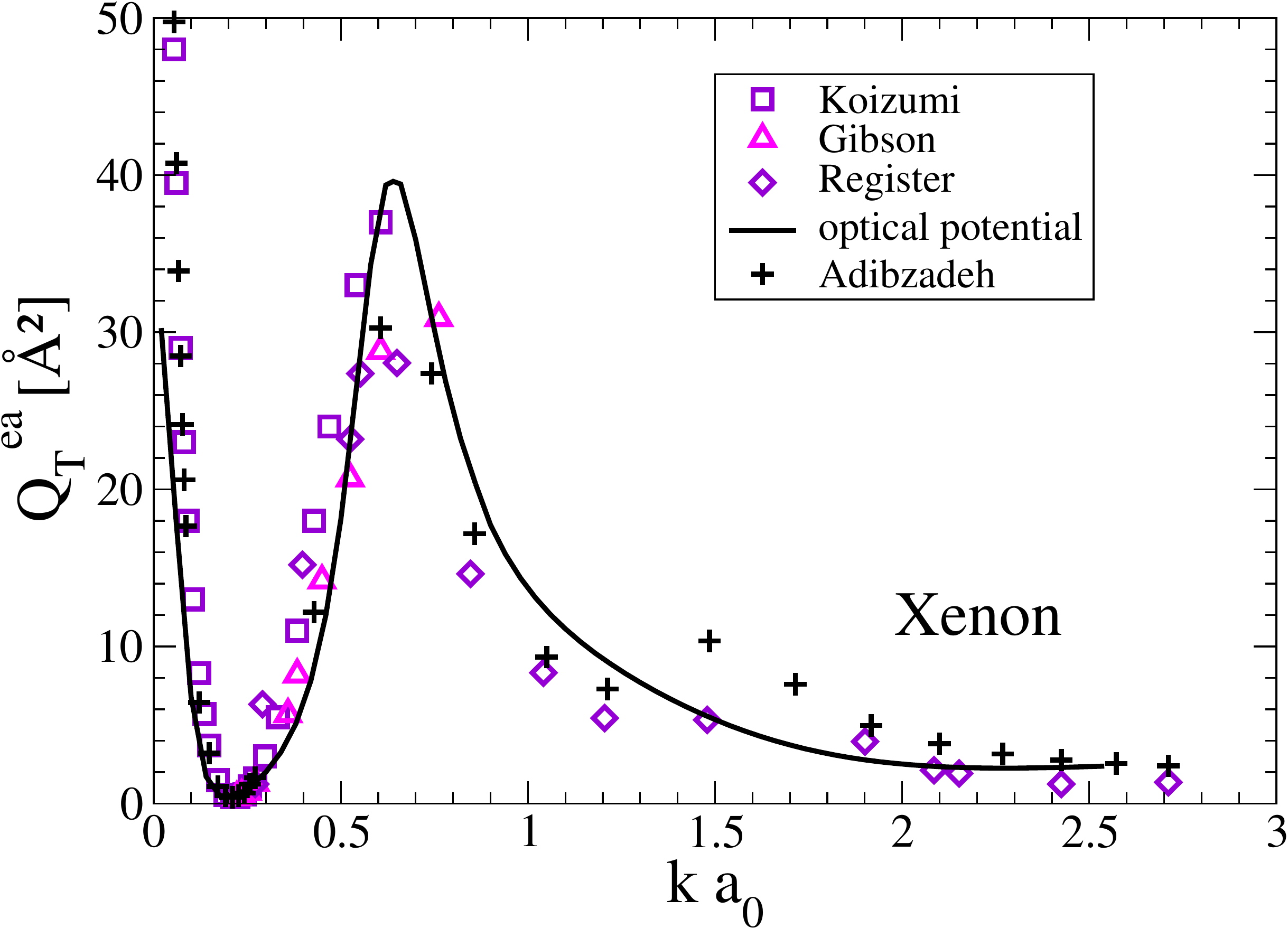}
\caption{(Color online) Momentum-transfer cross section for the noble gases. The calculations are performed using the optical potential Eq.~(\ref{eq:Vopt}) in the present work Eq.~(\ref{eq:VHF}),(\ref{eq:VPP}),(\ref{eq:Vex}) and (\ref{eq:Kex}) in comparison with the optical potential used in \cite{Adibzadeh05}. Also shown are experimental data \cite{Crompton70He,Milloy77He,Register80He,Williams79HeNeAr,Robertson72Ne,Register84Ne,Milloy77Ar,Gibson96Ar,Srivastava81ArKr,Panajo97Ar,England88Kr,Danjo88Kr,Koizumi86Xe,Gibson98Xe,Register86Xe}.}
\label{fig:QTall}
\end{center}
\end{figure}

\end{widetext}

In Fig.~\ref{fig:QTall} the results are compared with experimental data.
As a first result the form of the momentum-transfer cross section for all noble gases is in agreement with the experimental data. We conclude that the simple structure of helium as well as the specific structure of the other noble gases is well described by the optical potential. 
For low and intermediate energies the results are in good agreement with the experimental data. For instance, the shoulder for neon is reproduced as well as the Ramsauer minimum obtained for argon, krypton and xenon are reproduced. For large wavenumbers the discrepancies between the calculated and measured values increases for argon and krypton. 

To compare with other theoretical results calculations by Adibzadeh and Theodosiou \cite{Adibzadeh05} are given. The optical potential is used in a different way as in the present work.
For each element different exchange and polarization potentials are adjusted including four parameters.
The discrepancies with the experimental data are in the same order as the present results which only includes one adjusted parameter.
Adibzadeh and Theodosiou calculated for the heavier noble gases argon, kypton and xenon a smaller maximum in contrast to the proposed form for the optical potential. This difference can be explained by the different choice of the exchange and polarization term.

\subsection{Screening effects on the momentum-transfer cross section} \label{sec:sQT}
\begin{widetext}

\begin{figure}[htb] 
\begin{center}
 \includegraphics[width=0.4\linewidth,clip=true]{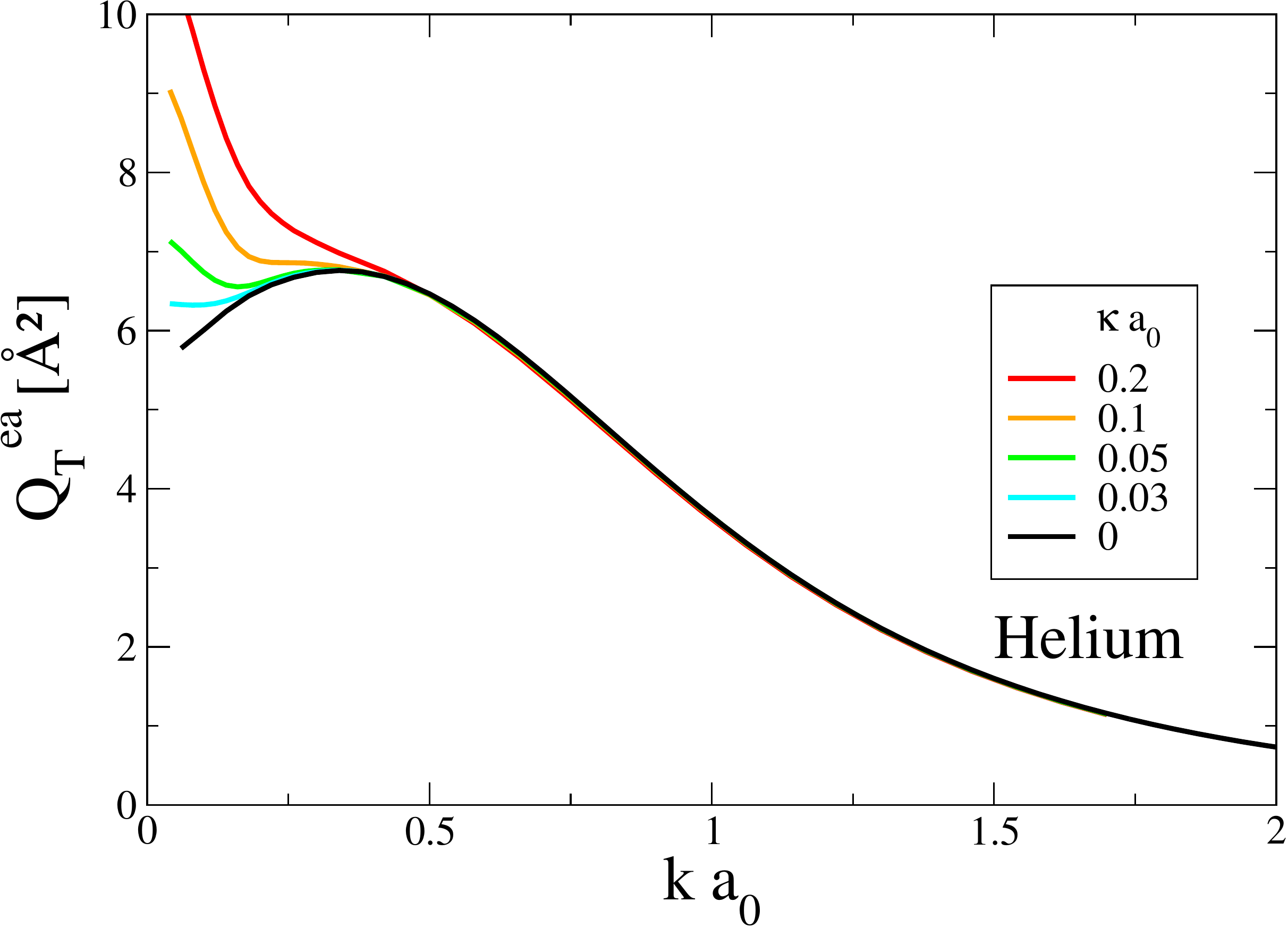} \,\,\,\,\,\,
 \includegraphics[width=0.4\linewidth,clip=true]{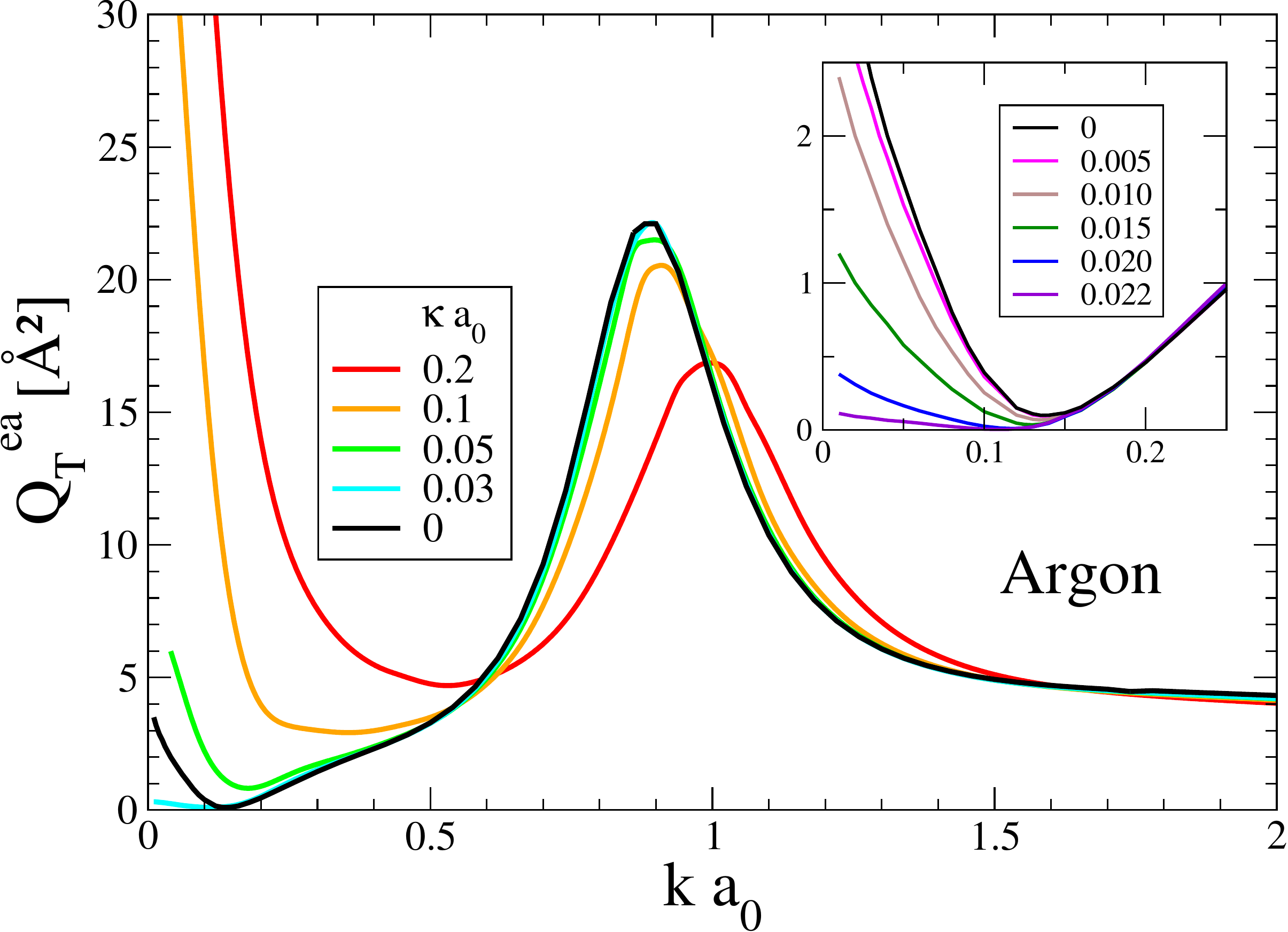}
\caption{(Color online) Screening effects on the momentum-transfer cross section for helium and argon at various screening parameters $\kappa$.}
\label{fig:sQTAr}
\end{center}
\end{figure}

\end{widetext}

Fig.~\ref{fig:sQTAr} shows the results of the momentum-transfer cross section using the screened optical potential for helium and argon.
As known from the $e$--$i$ interaction also in the case of $e$--$a$ collisions the plasma environment affects the transport of slower electrons more than for faster electrons.

First we consider argon.
For thin plasmas $\kappa a_0 < 0.025$ the momentum-transfer cross section decreases rapidly for slow electrons, the Ramsauer minimum is resolved at $\kappa a_0 \approx 0.025$. This is a consequence of the reduced zeroth scattering phase shift $\delta_0(k)$ for the repulsive screened optical potential at large distances. Around intermediate and high energies $ka_0 \geq 0.2$ discrepancies to the momentum-transfer cross section of isolated systems bescome small. The influence on the correlation functions is smaller than $1 \%$ for $T \approx 10^4 {\rm K}$. 
For high dense plasmas $\kappa a_0> 0.025$ the repulsive screened optical potential reduces the zeroth scattering phase shift $\delta_0(k)$ to values less than $3\pi$ so that the momentum-transfer cross section $Q_{\rm T}^{\rm ea}(k)$ increases. The new minimum which appears in the momentum-transfer cross section $Q_{\rm T}^{\rm ea}(k)$ is a result of higher phase shifts which is different to the Ramsauer effect. The minimum depth increases with the screening parameter $\kappa$ and the minimum position shifts to higher wave numbers. Furthermore the maximum position is also shifted to higher wavenumbers and the height decreases. The influence on the correlation functions is smaller than $2 \%$ for $T \approx 10^4 {\rm K}$. The effecte becomes more relevant for small temperaures.

We obtain similar results for helium. As a consequence of the first scattering phase shift $\delta_1$ a minimum appears in the momentum-transfer cross section. For higher densities the minimum depth increases and its position shifts to higher wave numbers as well as for argon. At $\kappa a_0 \approx 0.1$ the minimum changes to a saddle point and for higher screening parameters $\kappa$ to an inflection point. For low temperatures $T \approx 5 \, 000 {\rm K}$ the influence on the correlation functions is around $10 \%$.


\subsection{Correlation functions}
For partially ionized systems the composition is an ingrediant to calculate the influence of the electron-atom contribution.
The composition of the noble gases for a given temperature $T$ and mass density $\rho$ is calculated with Comptra04 \cite{Kuhlbrodt04}.

\begin{figure}[htb] 
\begin{center}
 \includegraphics[width=0.9\linewidth,clip=true]{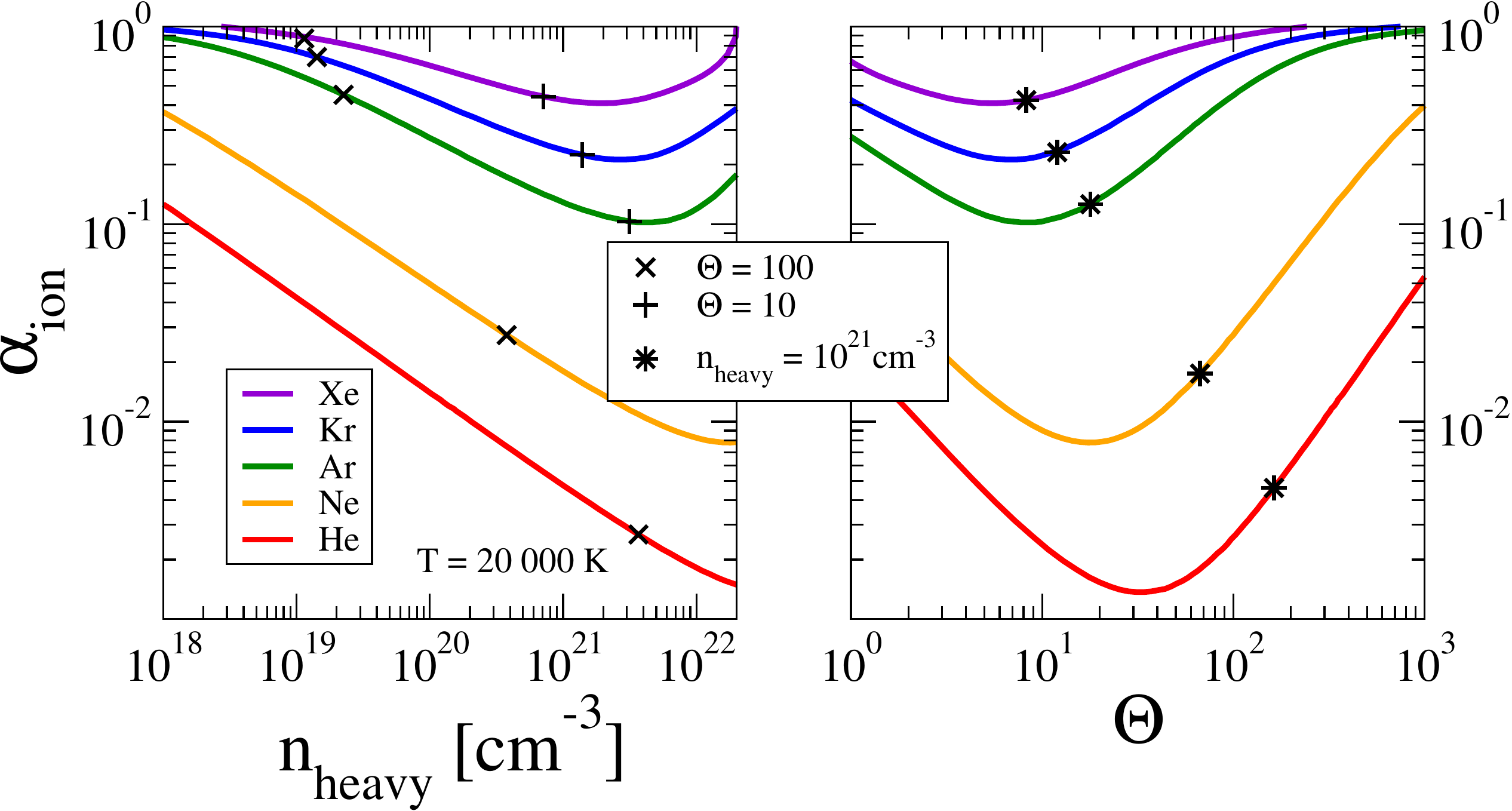}
\caption{(Color online) Ionization degree of the partially ionized noble gases at $T=20 \, 000 {\rm K}$.}
\label{fig:Comp}
\end{center}
\end{figure}

Fig.~\ref{fig:Comp} shows the ionization degree $\alpha_{\rm e}=n_{\rm e}/n_{\rm heavy}$ of the noble gases He, Ne, Ar, Kr and Xe at $T=20 \, 000 \, {\rm K}$ for heavy particle densities $n_{\rm heavy} = \rho N_{\rm A}/M=n_{\rm a}+n_{\rm i}$ between $10^{18} {\rm cm}^{-3} < n_{\rm heavy} < 10^{22} {\rm cm}^{-3}$ on the left side and in dependence on the degeneracy parameter $\Theta$ between $1 <\Theta < 1000$ on the right side. At low densities the ionization degree is $\approx 1$, we estimate a small contribution due to $e$--$a$ collisions for the electrical conductivity. For intermediate densities the ionization degree has a minimum and we estimate a dominant contribution by $e$--$a$ collisions. For high densities bound states are dissolved by the Mott effect the plasma becomes fully ionized $n_{\rm ea} \approx 0$. Because of Pauli blocking, also $e$--$e$ contributions can be neglected and only the $e$--$i$ contribution is important for the conductivity, the Ziman formula is applicable.

\begin{figure}[htb] 
\begin{center}
 \includegraphics[width=0.9\linewidth,clip=true]{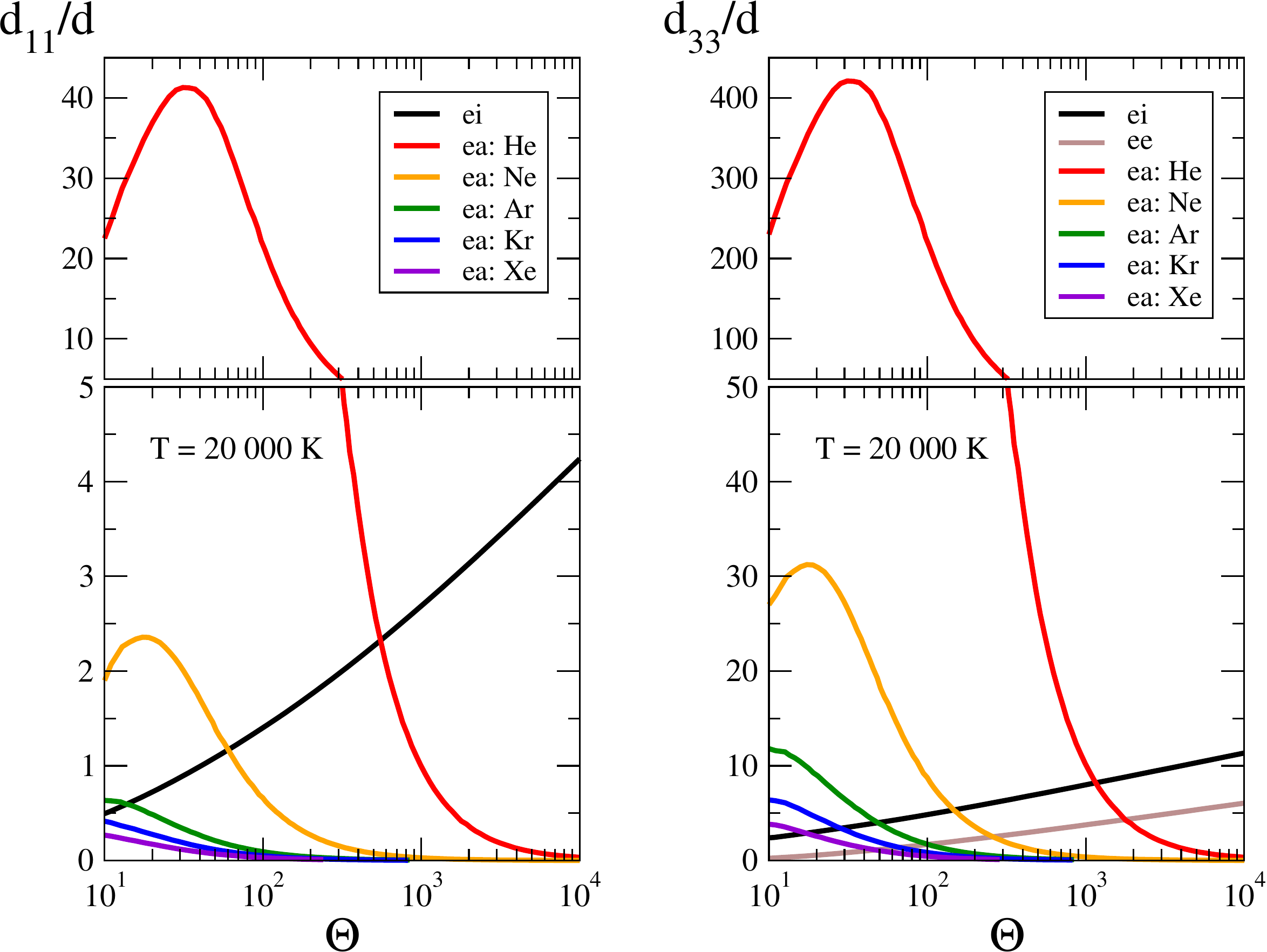}
\caption{(Color online) Correlation functions $d_{11}$ and $d_{33}$ in units of $d=\frac{4}{3}\sqrt{2\pi m \beta} [(n_{\rm e}e^2)/(4\pi\epsilon_0)]^2 \Omega_{\rm N}$ for the different considered scattering mechanisms $e$--$i$ (black), $e$--$e$ (brown), $e$--He (red), $e$--Ne (orange), $e$--Ar (green), $e$--Kr (blue) and $e$--Xe (violet) at $T = 20 \, 000 {\rm K}$. (Two scales.)}
\label{fig:CorrF}
\end{center}
\end{figure}

The different contributions for the correlation functions are shown in Fig.~\ref{fig:CorrF}. 
Despite the small ionization degrees the charged particle contributions yield the dominant parts for helium at $\Theta>600$ and neon at $\Theta > 60$ in $d_{11}$ (only $e$--$i$) and $d_{33}$ ($e$--$i$ and $e$--$e$) because of the weak $e$--$a$ interaction (weak $Q_{\rm T}^{\rm ea}$) in contrast to the charged particle interaction.
The $e$--$e$ contribution is explicitly given in $d_{33}$. For low densities the ratio between the $e$--$i$ and the $e$--$e$ contribution is $d_{33}^{\rm ei}/d_{33}^{\rm ee}>\sqrt{2}Z$.  

\subsection{Electrical conductivity for partially ionized noble gases}

\begin{widetext}

\begin{figure}[htb] 
\begin{center}
 \includegraphics[width=0.4\linewidth,clip=true]{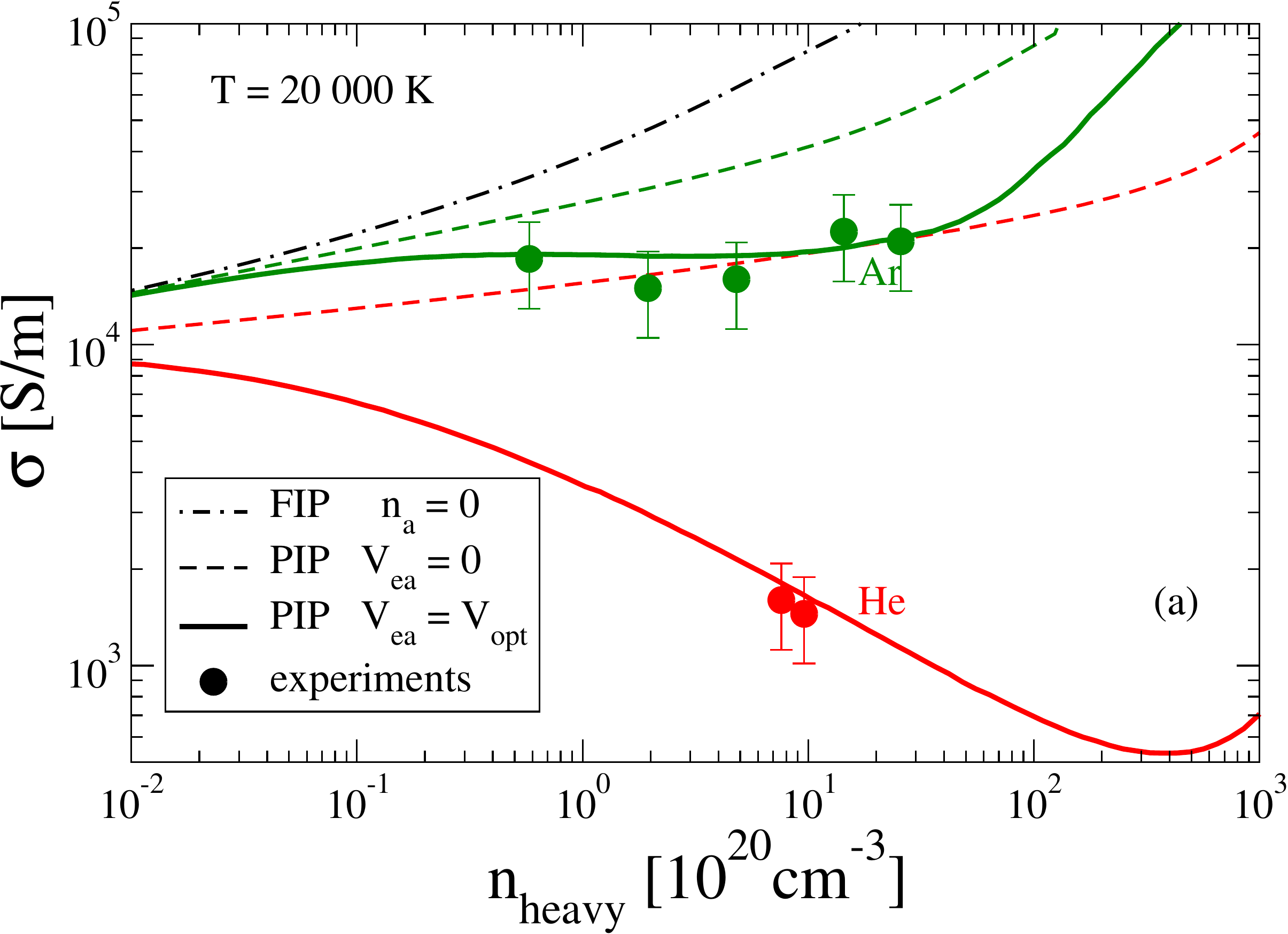}
 \includegraphics[width=0.4\linewidth,clip=true]{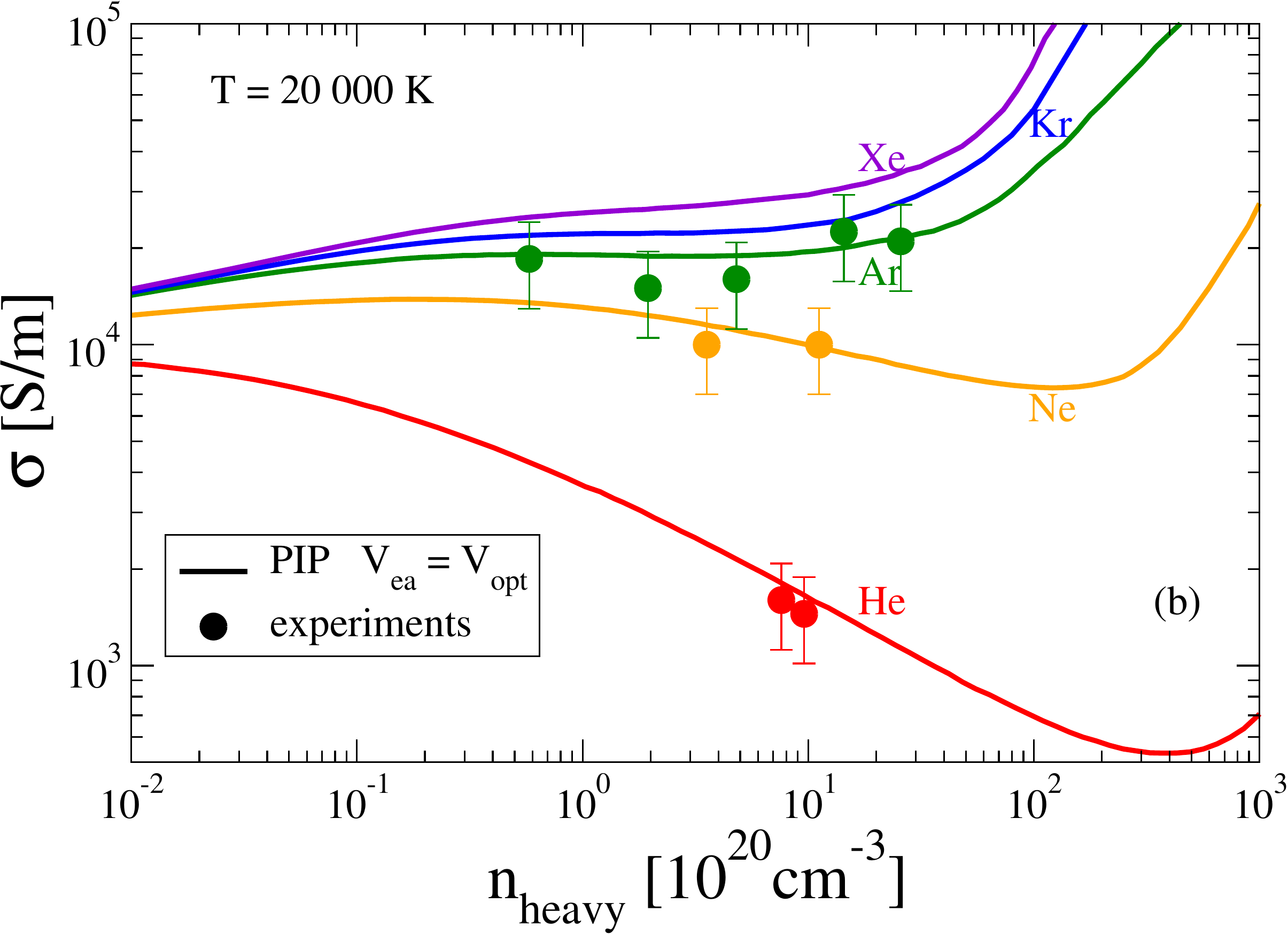}
 \includegraphics[width=0.4\linewidth,clip=true]{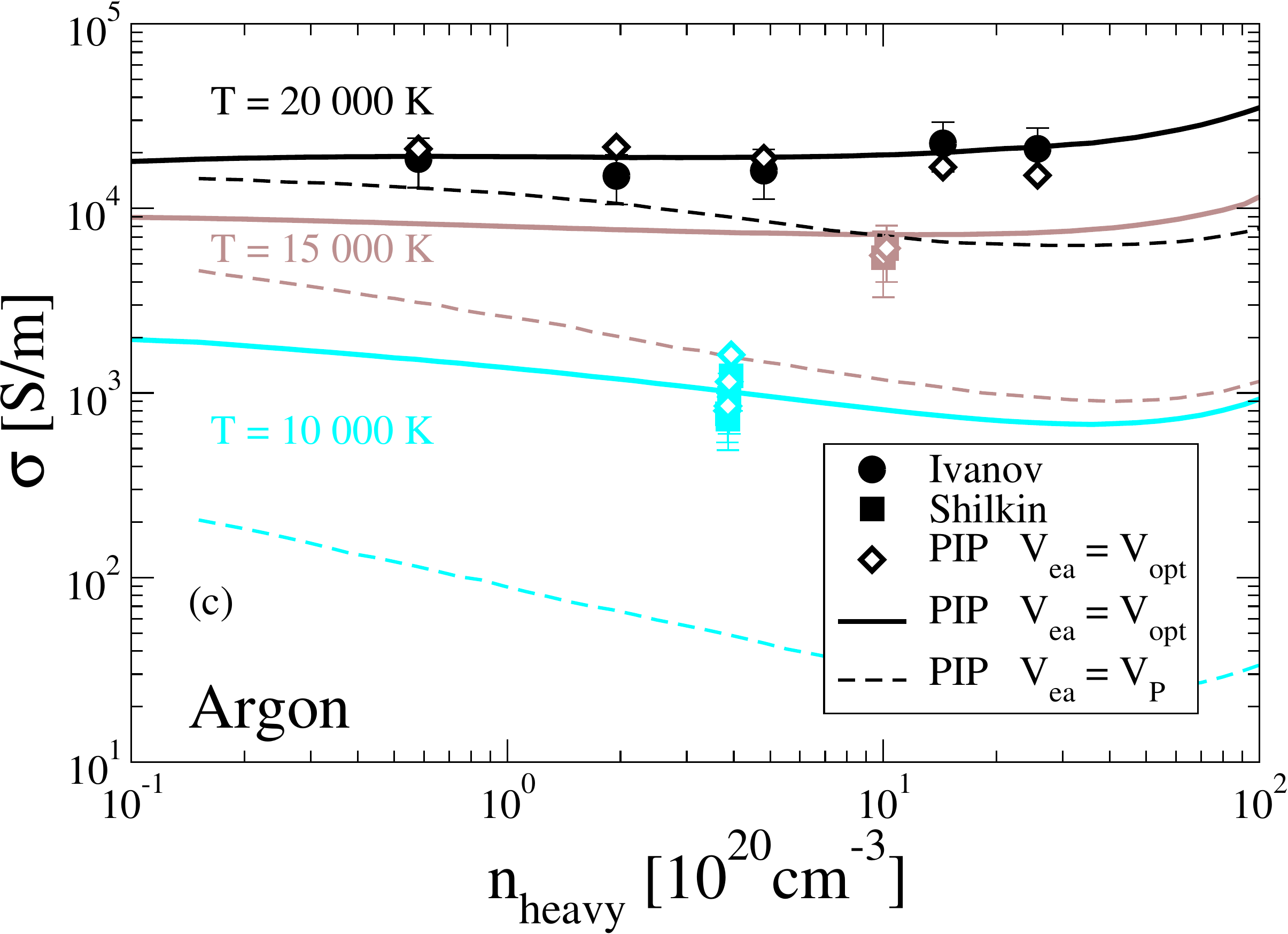}
 \includegraphics[width=0.4\linewidth,clip=true]{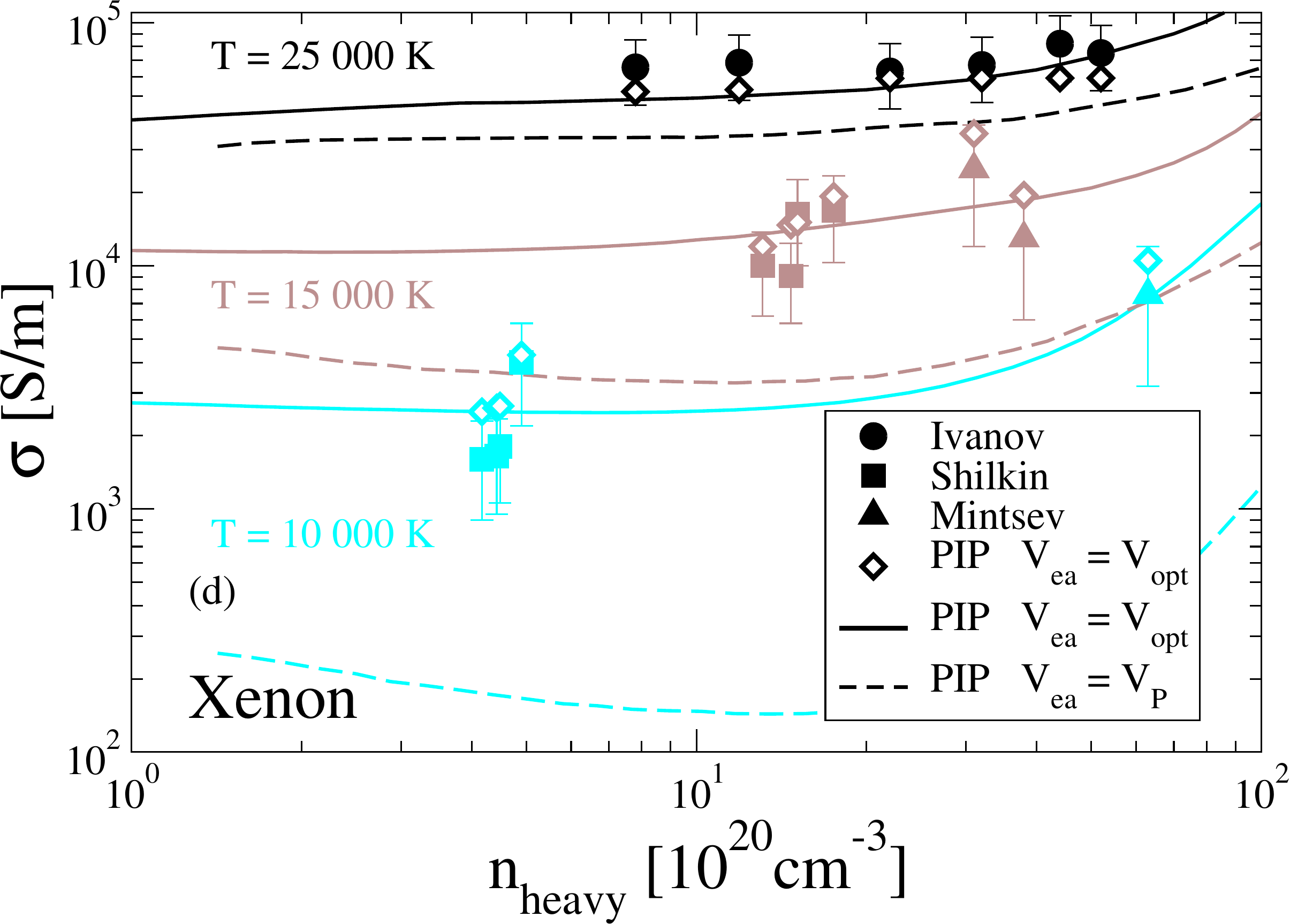}
\caption{Conductivity of partially ionized noble gases using the (screened) optical potential for the $e$--$a$ interaction (full lines) \\
(a): at $20 \, 000 {\rm K}$ for helium (red) and argon (green) in comparison with the fully ionized plasma model $n_{\rm a}=0$ (dashed-dotted) and an neglecting $e$--$a$ interaction $V_{\rm ea}=0$ (dashed) and experimental data \cite{Dudin98He,FortovHe,Ivanov76NeArXe} (circles); \\
(b): at $20 \, 000 {\rm K}$ for helium (red), neon (orange), argon (green), krypton (blue) and xenon (violet) in comparison with the experimental data \cite{Dudin98He,FortovHe,Ivanov76NeArXe} (circles); \\
(c): for argon at various temperatures $T=10 \, 000 {\rm K}$ (cyan), $15 \, 000 {\rm K}$ (brwon) and $20 \, 000 {\rm K}$ (black) in comparison with calculations using the polarization potential \cite{Kuhlbrodt05} (dashed) and experimental data \cite{Ivanov76NeArXe, Shilkin03ArXe} (filled circles, filled squares); \\
(d): for xenon at various temperatures $T=10 \, 000 {\rm K}$ (cyan), $15 \, 000 {\rm K}$ (brwon) and $25 \, 000 {\rm K}$ (black) in comparison with calculations using the polarization potential \cite{Kuhlbrodt05} (dashed) and experimental data \cite{Ivanov76NeArXe, Shilkin03ArXe, Mintsev79Xe} (filled circles, filled squares, filled triangles). }
\label{fig:PIPTra}
\end{center}
\end{figure}

\end{widetext}

Using the total momentum of electrons ${\bf P}_1$ and the heat current ${\bf P}_3$ as relevant observables the conductivity, Eq.~(\ref{eq:cond}), is calculated within the correlation functions $d_{11}$, $d_{13}$ and $d_{33}$ and the composition calculated with Comptra04. Theoretical results of the conductivity depending on the heave particle density $n_{\rm heavy}$ are compared with experimental data in Fig.~\ref{fig:PIPTra}.

In Fig.~\ref{fig:PIPTra} above (a,b) the strong reduction of the conductivity as the consequence of $e$-$a$ collisions is shown at $T\approx 20 \, 000 \, {\rm K}$. Corresponding to Fig.~\ref{fig:Comp} and \ref{fig:CorrF}, the reduction is stronger for the lighter elements. The characteristic minimum in the conductivity and a systematic behavior is observed for all noble gases. The experimental data for the conductivity are described by the PIP using the optical potential.
Similary good results are found by Adams {\it et al.} who used the experimental momentum-transfer cross sections describing $e$--$a$ collisions. 

In Fig.~\ref{fig:PIPTra} below (c,d) the conductivity depending on $n_{\rm heavy}$ is presented at various temperatures for argon and xenon. Experimental data are compared with calculations based on the optical potential, Eq.~(\ref{eq:sVopt}), and calculations performed by Kuhlbrodt {\it et al.} \cite{Kuhlbrodt05} based on the polarization potential for the description of $e$--$a$ collisions. 
Also for the lower temperatures $T=10 \, 000$ K the experimental data for the conductivity are described by the PIP using the optical potential. The qualitative behavior of the conductivity e.g. the characteristic minimum is also observed by Kuhlbrodt {\it et al.} \cite{Kuhlbrodt05} but the difference is up to two orders of magnitude, the polarization potential is not sufficient describing the $e$--$a$ collisions.

\section{Conclusion}
Within the LRT general expressions for plasma properties are obtained starting from a quantum-statistical approach. Introducing the chemical picture and the composition given by the ionization equilibrium the atoms (bound-states) are considered as a new species. Semi-empirical expressions for the effective potentials in particular for the $e$--$a$ interactions can be derived by first principle approaches. We introduce the optical potential to describe the $e$--$a$ interaction. A more systematic approach using the Green function method would give an interaction which is nonlocal and energy dependent. So the used optical potential is motivated by the perturbation theory up to second order which combines the Coulomb interaction which is relevant for the short distances and the polarization potential relevant for large distances. We used an exchange contribution determined by an effective field. This contribution is determiend by the condition that the characteristic scattering phase shifts of the non-local first order perturbation theory, the static-exchange approximation, are reproduced.

We show that a unique form of the optical potential can be introduced to describe the momentum-transfer cross section for all noble gases with high precision. Only one parameter $r_0$ is adjusted describing the $e$--$a$ scattering mechanism. Beyond the Born approximation specific effects e.g. the Ramsauer-Townsend minimum appear within a T matrix approach.

The optical potential model is extended to describe dense plasma environment in which screening effects arise.
The effects on the momentum-transfer cross sections are shown.
We show the Ramsauer-Townsend minimum is modified by plasma effects. Furthermore the electrical conductivity is calculated in a good agreement with experimental data.
We found that for plasma conditions $T \approx 1 {\rm eV}$ and $n_{\rm heavy} \approx 10^{21} {\rm cm}^{-3}$ the screening effects on the conductivity are smaller than $2 \%$. This explains why the calculations for the conductivity using the experimental momentum-transfer cross sections \cite{Adams07} are in a good agreement with the measured data. 

In general, to explore WDM but also ultra-cold gases the medium (plasma) effects may become more significant to describe transport properties. For future work improved screening effects (dynamical screening), the Pauli blocking, the static structure factor and degeneracy effects on the quantum mechanical T matrix can also be treated with our quantum-statistical approach.



	\section*{Acknowledgements}
The authors acknowledge support from the Deutsche Forschungsgemeinschaft (DFG) within the Collaborative Research Center SFB 652.

	\begin{appendix}
	
	\section{Screened Hartree-Fock potential} \label{AppA}
The Hartree-Fock potential between an electron and an atom describes the Coulomb interaction between the incoming electron with the core and the shell electrons given by
\begin{align}
 V_{\rm HF}(r) &= \frac{e^2}{4\pi \epsilon_0} \left[ -\frac{Z}{r} + \int \frac{1}{\bf |r-r'|}\rho(r') \, d^3{\bf r'} \right] \, .
\end{align}
For the plasma system the Coulomb interaction is screened. We replace the Coulomb interaction by a Debye potential
\begin{align}
 V_{\rm HF}^{\rm s}(r) &= \frac{e^2}{4\pi \epsilon_0} \left[ - \frac{Ze^{-\kappa r}}{r} + \int \frac{e^{-\kappa {\bf |r-r'|}}}{\bf |r-r'|}\rho(r') \, d^3{\bf r'} \right] \, .
\end{align}
The second term in the square brackets can be expressed by
\begin{align*}
2\pi \int\limits_0^{\infty} dr' \, r'^2 \rho(r') \int\limits_{-1}^{1} dz \, \frac{e^{-\kappa \sqrt{r^2+r'^2-2rr'z}}}{\sqrt{r^2+r'^2-2rr'z}}  \, .
\end{align*}
Substituting $y=\sqrt{r^2+r'^2-2rr'z}$ we obtain
\begin{align*}
 2\pi \int\limits_0^{\infty} dr' \, r'^2 \rho(r') \int\limits_{|r-r'|}^{r+r'} dy \, \frac{e^{-\kappa y}}{rr'}  \, .
\end{align*}
Performing the integral over $y$ the integral over $r'$ can be spitted, we obtain Eq.~(\ref{eq:sVHF}) which leads for $\kappa \rightarrow 0$ to the well known result Eq.~(\ref{eq:VHF}).

	\section{Asymptotic behavior for the screened Hartree-Fock potential} \label{AppB}

Expanding the exponential functions in the integrand of $I_1$ and $I_2$ we obtain:
\begin{align}
 I_1 &= \frac{e^{-\kappa r}}{2\kappa r} \int\limits_0^{r} \frac{\rho(r_1)}{r_1} \left\{1+ \kappa r_1 + \frac{\kappa^2 r_1^2}{2} +...  \right\} \, dr_1 \, , \\
 I_2 &= - \frac{e^{-\kappa r}}{2\kappa r}  \int\limits_0^{\infty} \frac{\rho(r_1)}{r_1} \left\{ 1 - \kappa r_1 + \frac{\kappa^2 r_1^2}{2} -+... \right\}  \, dr_1 \, , \\
 I_1+I_2 &> \frac{Ze^{-\kappa r}}{r} - \frac{1}{2\kappa r} e^{-\kappa r} \int\limits_r^{\infty} \frac{\rho(r_1)}{r_1} \left\{ 1 + \kappa r_1 + \kappa^2 r_1^2 \right\} \, . \label{eq:I1I2}
\end{align}
For large distances $r \rightarrow \infty$, $I_3$ is bigger than the last term in Eq.~(\ref{eq:I1I2}),
so that $V_{\rm HF}(r)$ becomes repulsive. 
%

Expanding the exponential functions for $I_1$ and $I_2$ up to the third order $(\kappa r_1)^3$ we find
\begin{align}
\begin{split}
 V_{\rm HF}^{\rm s}(r)&= \frac{e^2}{4\pi \epsilon_0}\frac{e^{-\kappa r}}{\kappa r} \int\limits_0^{\infty} \frac{\rho(r_1)}{r_1} \left\{\frac{\kappa^3 r_1^3}{3!} + \frac{\kappa^5 r_1^5}{5!} +  ...  \right\} \, dr_1 \\
& \quad- \frac{e^2}{4\pi \epsilon_0} \frac{1}{\kappa r} \int\limits_r^{\infty} \frac{\rho(r_1)}{r_1} \sinh\left[\kappa (r_1 -r)\right] \, dr_1 \, .
\end{split}
\end{align}
For $0 < \kappa a_0 < 1$ at the dominant contribution in the first term is in order to ${\cal O} (\kappa^2 e^{-\kappa r}/r)$. The last term decreases faster for large distances $r \rightarrow \infty$, we obtain
\begin{align}
 V_{\rm HF}^{\rm s}(r) &= \frac{Z e^2}{4\pi \epsilon_0 } \frac{e^{-\kappa r}}{r} \, \sum_{k=0}^{\infty} (\kappa a_0)^{k+2} {\cal C}_k \, , \\
 {\cal C}_k &= \frac{Z^{-1}}{(2k+3)!} \int\limits_0^{\infty} \left( \frac{r_1}{a_0}\right)^{2k+2} \rho(r_1) \, dr_1 \, .
\end{align}
The numerical values for the coefficients ${\cal C}_k$ are listed in Tab.~\ref{tab:coef}.
Because of the fast convergence only ${\cal C}_0$ is necessary at an inverse screening length $0<\kappa a_0 < 1$, Eq.~(\ref{eq:asVHF}) is veryfied.

\begin{center}
\begin{table}[h]
 \begin{tabular}{ | c | c c c c c |}
\hline
$k$ & ${\cal C}_k^{\rm He}$ & ${\cal C}_k^{\rm Ne}$ & ${\cal C}_k^{\rm Ar}$ & ${\cal C}_k^{\rm Kr}$ & ${\cal C}_k^{\rm Xe}$  \\ \hline
0 & 0.1974 & 0.1563 & 0.2411 & 0.1830 & 0.1933  \\
1 & 0.0324 & 0.0228 & 0.0672 & 0.0568 & 0.0727 \\
2 & 0.0050 & 0.0033 & 0.0170 & 0.0165 & 0.0262 \\
3 & 0.0007 & 0.0005 & 0.0044 & 0.0044 & 0.0088 \\
4 & 0.0001 & 0.0001 & 0.0013 & 0.0011 & 0.0028 \\
\hline
 \end{tabular}
 \caption{Coefficients ${\cal C}_k$ for the noble gases.}
\label{tab:coef}
\end{table}
\end{center}

	\end{appendix}

%

\end{document}